\documentclass{aa}  

\usepackage{graphicx}
\usepackage{xcolor}
\usepackage{booktabs}
\usepackage{supertabular}
\usepackage[version=4]{mhchem}
\usepackage{ulem}
\usepackage{longtable}
\usepackage{caption}
\usepackage{sidecap}
\usepackage{float}
\usepackage{placeins}
\usepackage{rotating}
\usepackage{txfonts}
\usepackage{hyperref}
\hypersetup{
    colorlinks=true,
    linkcolor= blue,
    filecolor=magenta,      
    urlcolor=blue,
    citecolor= blue,
}%
\begin{document}

   \title{A deep search for large complex organic species toward IRAS16293-2422 B at 3\,mm with ALMA}

   \author{P. Nazari
          \inst{1}
          \and
          J. S. Y. Cheung\inst{2}
          \and
          J. Ferrer Asensio\inst{3}
          \and
          N. M. Murillo\inst{4, 5}
          \and
          E. F. van Dishoeck\inst{1,3}
          \and
          J. K. J{\o}rgensen\inst{6}
          \and 
          T. L. Bourke\inst{7}
          \and
          K. -J. Chuang\inst{8}
          \and
          M. N. Drozdovskaya\inst{9}
          \and
          G. Fedoseev\inst{10}
          \and
          R. T. Garrod\inst{11}
          \and
          S. Ioppolo\inst{12}
          \and
          H. Linnartz\inst{8}
          \and
          B. A. McGuire\inst{2, 13}
          \and
          H. S. P. M\"{u}ller\inst{14}
          \and
          D. Qasim\inst{15}
          \and
          S. F. Wampfler\inst{9}  
          }

   \institute{Leiden Observatory, Leiden University, P.O. Box 9513, 2300 RA Leiden, the Netherlands\\ 
        \email{nazari@strw.leidenuniv.nl}
         \and
         Department of Chemistry, Massachusetts Institute of Technology, Cambridge, MA 02139, USA
         \and
         Max-Planck-Institut f\"{u}r extraterrestrische Physik, Giessenbachstr. 1, 85748 Garching, Germany
         \and
         Instituto de Astronom\'{i}a, Universidad Nacional Aut\'{o}noma de M\'{e}xico, AP106, Ensenada CP 22830, B. C., M\'{e}xico
         \and
         Star and Planet Formation Laboratory, RIKEN Cluster for Pioneering Research, Wako, Saitama 351-0198, Japan
         \and
         Niels Bohr Institute, University of Copenhagen, {\O}ster Voldgade 5–7, 1350 Copenhagen K., Denmark
         \and
         SKA Organization, Jodrell Bank Observatory, Lower Withington, Macclesfield, Cheshire SK11 9FT, UK
         \and
         Laboratory for Astrophysics, Leiden Observatory, Leiden University, P.O. Box 9513, 2300 RA Leiden, the Netherlands
         \and
         Center for Space and Habitability, Universit\"{a}t Bern, Gesellschaftsstrasse 6, CH-3012 Bern, Switzerland
         \and
         INAF - Osservatorio Astrofisico di Catania, via Santa Sofia 78, 95123 Catania
         \and
         Departments of Astronomy and Chemistry, University of Virginia, Charlottesville, VA 22904, USA
         \and
         Center for Interstellar Catalysis, Department of Physics and Astronomy, Aarhus University, Ny Munkegade 120, Aarhus C 8000, Denmark
         \and
         National Radio Astronomy Observatory, Charlottesville, VA 22903, USA
         \and
         Astrophysik/I. Physikalisches Institut, Universit\"{a}t zu K\"{o}ln, Z\"{u}lpicher Str. 77, 50937 K\"{o}ln, Germany
         \and
         Southwest Research Institute, San Antonio, TX 78238, USA
         }

   \date{Received 30 August 2023 / Accepted 13 December 2023}

 
  \abstract
   {Complex organic molecules (COMs) have been detected ubiquitously in protostellar systems. However, at shorter wavelengths (${\sim}0.8$\,mm) it is generally more difficult to detect larger molecules than at longer wavelengths (${\sim}3$\,mm) because of the increase of millimeter dust opacity, line confusion, and unfavorable partition function.}
   {We aim to search for large molecules ($> 8$ atoms) in the Atacama Large Millimeter/submillimeter Array (ALMA) Band 3 spectrum of IRAS 16293-2422 B. In particular, the goal is to quantify the usability of ALMA Band 3 for molecular line surveys in comparison to similar studies at shorter wavelengths.}
   {We use deep ALMA Band 3 observations of IRAS 16293-2422 B to search for more than 70 molecules and identify as many lines as possible in the spectrum. The spectral settings were set to specifically target three-carbon species such as i- and n-propanol and glycerol, the next step after glycolaldehyde and ethylene glycol in the hydrogenation of CO. We then derive the column densities and excitation temperatures of the detected species and compare the ratios with respect to methanol between Band 3 (${\sim}3$\,mm) and Band 7 (${\sim}1$\,mm, Protostellar Interferometric Line Survey) observations of this source to examine the effect of dust optical depth.}
   {We identify lines of 31 molecules including many oxygen-bearing COMs such as CH$_3$OH, CH$_2$OHCHO, CH$_3$CH$_2$OH, c-C$_2$H$_4$O and a few nitrogen- and sulfur-bearing ones such as HOCH$_2$CN and CH$_3$SH. The largest detected molecules are gGg-(CH$_2$OH)$_2$ and CH$_3$COCH$_3$. We do not detect glycerol or i- and n-propanol but provide upper limits for them which are in line with previous laboratory and observational studies. The line density in Band 3 is only ${\sim}2.5$ times lower in frequency space than in Band 7. From the detected lines in Band 3 at a $\gtrsim 6\sigma$ level, ${\sim}25-30\%$ of them could not be identified indicating the need for more laboratory data of rotational spectra. We find similar column densities and column density ratios of COMs (within a factor ${\sim}2$) between Band 3 and Band 7.}
   {The effect of dust optical depth for IRAS 16293-2422 B at an off-source location on column densities and column density ratios is minimal. Moreover, for warm protostars, long wavelength spectra (${\sim}3$\,mm) are not only crowded and complex, but also take significantly longer integration times than shorter wavelength observations (${\sim}0.8$\,mm) to reach the same sensitivity limit. The 3\,mm search has not yet resulted in detection of larger and more complex molecules in warm sources. A full deep ALMA Band $2-3$ (i.e., ${\sim}3-4$\,mm wavelengths) survey is needed to assess whether low frequency data have the potential to reveal more complex molecules in warm sources.}

   \keywords{Astrochemistry --
                Stars: low-mass --
                Stars: protostars --
                ISM: abundances --
                ISM: molecules
               }

   \maketitle
%

\section{Introduction}

In the interstellar medium (ISM) complex organic molecules (COMs), defined as species with at least 6 atoms containing carbon (\citealt{Herbst2009}), are particularly prominent in the protostellar phase. Although other phases of star formation such as the prestellar phase (e.g., \citealt{Bacmann2012}; \citealt{JimenezSerra2016}; \citealt{McGuire2020}; \citealt{Scibelli2020}) and the later protoplanetary disk phase (\citealt{Oberg2015}; 
\citealt{Walsh2016}; \citealt{Booth2021}; \citealt{Brunken2022}) show detections of these species, COMs are easier detectable in the line-rich protostellar envelopes due to their higher temperatures (e.g., \citealt{Blake1987}; \citealt{Belloche2013}; \citealt{Bergner2017}; \citealt{vanGelder2020}; \citealt{Nazari2021}; \citealt{Yang2021}; \citealt{McGuire2022}; \citealt{Bianchi2022}; \citealt{Hsu2022}).

Many COMs, including species with more than 8 atoms, are expected to form in ices under laboratory conditions (ethanol (CH$_3$CH$_2$OH), \citealt{Oberg2009}; \citealt{Chuang2020}; \citealt{Fedoseev2022}; aminomethanol (NH$_2$CH$_2$OH), \citealt{Theule2013}; glycerol (HOCH$_2$CH(OH)CH$_2$OH), \citealt{Fedoseev2017}; 1-propanol (CH$_3$CH$_2$CH$_2$OH), \citealt{Qasim2019, Qasim2019_Cocktail}; glycine (NH$_2$CH$_2$COOH), \citealt{Ioppolo2021}). However, there is a lot of debate regarding the ice or gas-phase formation of particular COMs (e.g., \citealt{Ceccarelli2022}). Two examples of these species are formamide (NH$_2$CHO) and acetaldehyde (CH$_3$CHO) for which both gas and ice formation pathways are suggested (e.g., \citealt{Jones2011}; \citealt{Barone2015}; \citealt{Vazart2020}; \citealt{Chuang2020, Chuang2021, Chuang2022}; \citealt{Fedoseev2022}; \citealt{Garrod2022}). To obtain clues to the formation mechanism of COMs from an observational perspective, it is possible to search for the solid state signatures of COMs in ices (\citealt{Schutte1999}; \citealt{Oberg2011}) with telescopes such as the \textit{James Webb Space Telescope} (\citealt{Yang2022}; \citealt{McClure2023}; Rocha et al. in prep.) using laboratory spectra available from for example, Leiden Ice Database for Astrochemistry (\citealt{Rocha2022}) and to examine the gas-phase correlations between different COMs in large samples of sources (\citealt{Belloche2020}; \citealt{Coletta2020}; \citealt{Jorgensen2020}; \citealt{Nazari2022ALMAGAL}; \citealt{Taniguchi2023}; \citealt{Chen2023}). However, observations and models show that physical effects such as source structure or dust optical depth could affect the snowline locations, gas-phase emission, and column density correlations of simple and complex molecules (\citealt{Jorgensen2002}; \citealt{Persson2016}; \citealt{DeSimone2020}; \citealt{Nazari2022}; \citealt{Murillo2022models}; \citealt{Nazari2023Massive, Nazari2023Scatter}). Hence, interpretations of COM formation routes based on gas-phase observations can be affected by these physical factors. For example, an anti-correlation between column densities of two species (or a large scatter in the column density ratios) could have a physical origin rather than a chemical origin.

Among the low-mass protostars, IRAS16293-2422 (IRAS16293 hereafter) triple protostellar system (\citealt{Wootten1989}; \citealt{Maureira2020}) is one of the closest protostars, and one of the richest and most well-studied objects from a chemical perspective. The first detection of methanol (CH$_3$OH; the simplest COM) toward a low-mass protostar was made by \cite{Ewine1995} for this system. Since then, its chemistry and in particular COMs in this system have been studied in more detail (\citealt{Cazaux2003}; \citealt{Butner2007}; \citealt{Bisschop2008}; \citealt{Ceccarelli2010}; \citealt{Jorgensen2011}; \citealt{Jorgensen2012}; \citealt{Kahane2013}; \citealt{Jaber2014}). More recently, the Protostellar Interferometric Line Survey (PILS; \citealt{Jorgensen2016}) studied IRAS16293 in Band 7 (${\sim}329.147-362.896$\,GHz) of the Atacama Large Millimeter/submillimeter Array (ALMA). This survey detected many COMs for the first time in the interstellar medium adding further information on complexity in space (\citealt{Coutens2016}; \citealt{Lykke2017}; \citealt{Calcutt2018First}; \citealt{Jorgensen2018}; \citealt{Manigand2019}; \citealt{Manigand2021}; \citealt{Coutens2022}). 

However, a limitation of higher-frequency observations of ALMA (${\sim}330$\,GHz) is the higher degree of line blending. The reason is that for observations probing the same gas the line width, although constant in velocity space, increases in frequency space (\citealt{Jorgensen2020}). Therefore, the detection of larger COMs with more than 8 atoms, that have relatively weak lines, is expected to be easier at lower frequencies. Moreover, the heavier molecules have their Boltzmann distribution peak at lower frequencies than the lighter molecules at the same excitation temperature, thus their lines are stronger at lower frequencies and they are easier detected at long wavelength observations. On the other hand, the Boltzmann peak moves to the higher frequencies for higher temperatures (see Fig. 2 of \citealt{Herbst2009}). Therefore, if the large molecules are thermally sublimated in the inner hot regions around the protostar (tracing warm or hot regions) it will be difficult to observe these larger species even at low frequencies (although favored), unless the data is sensitive enough, which can be achieved at the expense of the angular resolution and longer integration times. The combination of the increase in line width in frequency space and the Boltzmann distribution peak of the lighter molecules being at higher frequencies is also that the spectra are more crowded and thus, it is more difficult to detect large molecules at high frequencies.

Moreover, dust optical depth effects could be an important issue at higher-frequency ALMA Bands 6 (${\sim}240$\,GHz) and 7 (${\sim}330$\,GHz). This is because of the larger dust opacity of ${\sim}1$\,mm-sized grains at shorter wavelengths (i.e, higher frequencies). For example, \cite{Lopez2017} found that NGC 1333 IRAS 4A1 in Perseus does not host any COMs when searched for with ALMA and the Plateau de Bure Interferometer. However, later \cite{DeSimone2020} detected methanol around this source at longer wavelengths with the Very Large Array (VLA). Another example is the ring-shaped structure of methanol around the dust continuum in the massive protostellar system 693050 (also known as G301.1364-00.2249) observed by \cite{vanGelder2022} at ${\sim}220$\,GHz with ALMA, which indicates dust attenuation on-source.

In this work, we use the deep Band 3 ALMA observations of IRAS16293 B to search for larger species (> 8 atoms). This data set was specifically optimized to hunt for molecules such as glycerol and i- and n-propanol. Although the main aim is to specifically hunt for large molecules, we identify as many molecules as possible from the Cologne Database for Molecular Spectroscopy (CDMS; \citealt{Muller2001}; \citealt{Muller2005}) and the Jet Propulsion Laboratory database (JPL; \citealt{Pickett1998}) in our data. We fit the spectrum to derive the column densities and excitation temperatures of the detected species and compare our results with those of PILS in Band 7. In particular, we examine whether dust attenuation is important for column densities and their ratios with respect to methanol (typically used as a reference species).

The paper is structured such that the observations are explained in Sect. \ref{sec:meth_obs}. The results including the detected species and their column densities and excitation temperatures are given in Sect. \ref{sec:results}. We discuss our findings, in particular the comparison with the PILS results in Sect. \ref{sec:discussion}. Finally, we present our conclusions in Sect. \ref{sec:conclusions}.  

\section{Observations and methods}
\label{sec:meth_obs}
\subsection{Data}

\begin{figure*}
    \centering
    \includegraphics[width=17cm]{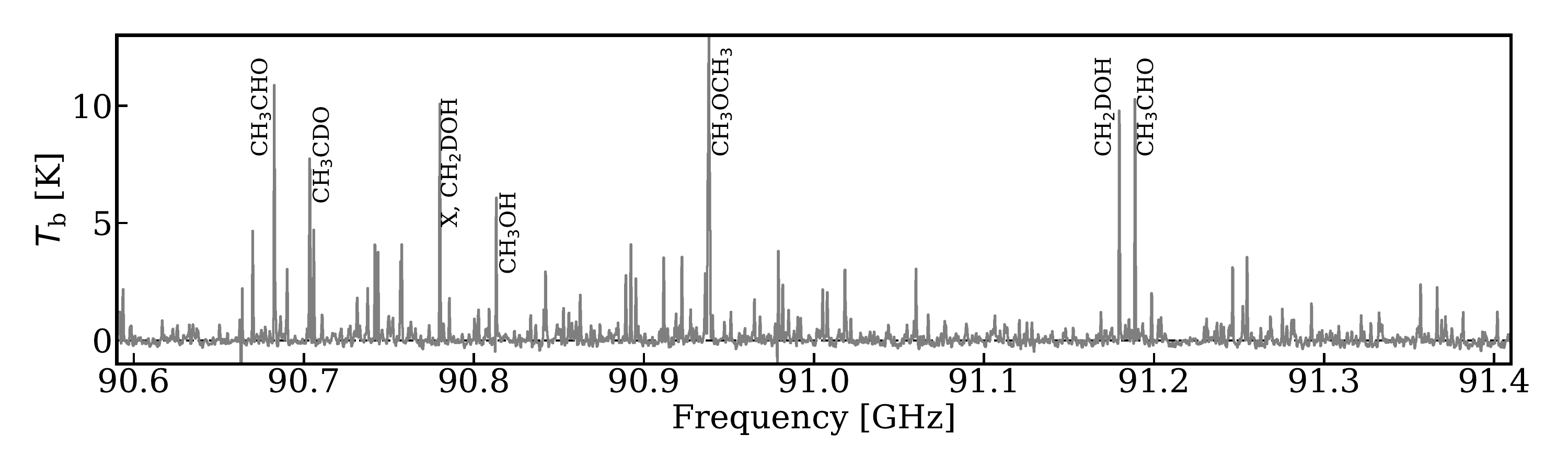}
    \caption{The spectrum of IRAS16293 B Band 3 data for the continuum window. A few strong lines are highlighted, for the full line identification and fitting see Fig. \ref{fig:model_spec}. `X' indicates a line that is not yet identified and is not (fully) fitted with the considered molecules in this work. The spectrum of IRAS16293 B is line rich even at ALMA Band 3 wavelengths (${\sim}3$\,mm).} 
    \label{fig:overview}
\end{figure*}

This paper uses the data of IRAS16293 B taken with ALMA in Band 3 (project code: 2017.1.00518.S; PI: E.~F. van Dishoeck). For more information on the data calibration, reduction and imaging, as well as first results on gas accretion flow in the IRAS16293 system see \cite{Murillo2022}, here we give a brief description of the data. We used the data with configuration C43-4 which were calibrated with the CASA pipeline (\citealt{McMullin2007}) and then self-calibrated with CASA. Continuum subtraction was done on the self-calibrated image datacubes using STATCONT (\citealt{Sanchez2018}). 

The data have an angular resolution of ${\sim} 0.9\arcsec \times 0.7\arcsec$ which is larger than that of PILS (${\sim}0.5\arcsec$). The frequency ranges covered were ${\sim}90.59-91.41$\,GHz (continuum), ${\sim}93.14-93.20$\,GHz and ${\sim}103.28-103.52$\,GHz. After conversion of the flux to brightness temperature the rms found from line-free regions in each spectral window is ${\sim}0.1$\,K, ${\sim}0.3$\,K and $0.2$\,K (comparable to PILS), respectively, following 171.7 minutes on-source integration time. The spectral resolution was 0.805\,km\,s$^{-1}$, 0.049\,km\,s$^{-1}$, and 0.088\,km\,s$^{-1}$. Given that the lines have full width at half maximum (FWHM) of ${\sim}1-2$\,km\,s$^{-1}$, some lines are not fully resolved in the continuum window. The frequency range was optimized to cover transitions of larger species such as glycerol, i- and n-propanol. It also serendipitously covered transitions from other oxygen- and nitrogen-bearing complex organic molecules including glycolaldehyde (CH$_{2}$OHCHO), propanal (g-C$_2$H$_5$CHO) and glycolonitrile (HOCH$_2$CN). The spectrum of IRAS16293 B (Fig. \ref{fig:overview}) was extracted from the same $0.5\arcsec$ offset position from the continuum peak of B that was used in many of the works from ALMA-PILS ($\alpha_{\rm J2000} = 16^{\rm h} 32^{\rm m} 22^{\rm s}.58$ and $\delta_{J2000} = -24^{\circ} 28\arcmin 32.8\arcsec$, \citealt{Coutens2016}; \citealt{Jorgensen2018}; \citealt{Calcutt2018}). Maps of a few common species tracing cold gas structures are presented in \cite{Murillo2022}.

\subsection{Spectral modeling}
\label{sec:spec_modeling}
In this work, we identified as many molecules as possible using the CASSIS\footnote{\url{http://cassis.irap.omp.eu/}} spectral analysis tool (\citealt{Vastel2015}). We consider a molecule as detected if it has at least three lines at a peak of $\geq3\sigma$ level. We also report column densities of the molecules that have one or two detected lines without any over prediction of the emission in the spectrum. These column densities are only tentative and not as robust as those found for molecules with many detected lines because of the limited frequency coverage but should be reliable for the most abundant and common molecules such as $^{13}$CH$_3$OH. Using CASSIS, we fitted for the identified molecules. In the fitting procedure, the spectrum was considered as a whole; and all the lines of a molecule were fitted simultaneously assuming local thermodynamic equilibrium (LTE) conditions. The fitting followed a similar procedure as the fit-by-eye method used in \cite{Nazari2022ALMAGAL}. This includes varying the column densities ($N$) and excitation temperatures ($T_{\rm ex}$) in the LTE models to produce a synthetic spectrum that matches the data (for more information on this process see below). The line lists were taken from the CDMS (\citealt{Muller2001}; \citealt{Muller2005}) or the JPL database (\citealt{Pickett1998}). Appendix \ref{sec:spec_data} gives more information on the spectroscopic studies used for our assignments and Table \ref{tab:lines} presents the transitions covered in the data.

In the fitting process, the FWHM was first fitted for the unblended line(s) of each molecule. Then it was fixed to that value when the column density and excitation temperature were being determined. The FWHM for all considered species is ${\sim}1.8\pm 0.5$\,km\,s$^{-1}$, except for (formic acid) t-HCOOH which is slightly lower (i.e., ${\sim}1$\,km\,s$^{-1}$). The typical FWHM here is ${\sim}1.5-2$ times larger than the typical FWHM found in PILS (${\sim}1$\,km\,s$^{-1}$). However, the spectral resolution for the continuum spectral window (i.e., the major spectral window where most lines lie; Fig. \ref{fig:model_spec}) is ${\sim}0.8$\,km\,s$^{-1}$ in this work. This implies that not all lines are spectrally resolved. Therefore, the higher FWHM measured here could be due to this low spectral resolution. The excitation temperature was only fitted if there were enough ($\geq 2$) detected lines of a molecule with a range of $E_{\rm up}$ (e.g., ${\sim}100-400$\,K). The typical uncertainty on the excitation temperature is ${\sim}\pm50$\,K. If not enough lines with a range of $E_{\rm up}$ were detected, the temperature is fixed to 100\,K, which is similar to excitation temperatures assumed or found for many species in PILS (\citealt{Lykke2017}; \citealt{Calcutt2018}; \citealt{Jorgensen2018}). The exceptions to this rule were t-HCOOH and NH$_2$CHO. Only one line was detected for each of these two molecules and their temperatures were fixed to 300\,K to be consistent with PILS (\citealt{Coutens2016}; \citealt{Jorgensen2018}). Moreover, for the isotopologues for which a determination of the excitation temperature was not possible, the temperature was fixed to that of the other isotopologues with determined $T_{\rm ex}$. Deuterated methanol, CH$_2$DOH, has three detected lines, with upper energy levels of ${\sim}10$\,K, ${\sim}100$\,K and ${\sim}400$\,K. It was not possible to fit all three lines of this molecule with a single excitation temperature. Therefore, we adopted the same temperature as the methanol isotopologue that has the most number of lines detected (i.e., CHD$_2$OH). However, this temperature fits the ${\sim}100$\,K line better than the other two. For aGg$^{\prime}$-(CH$_2$OH)$_2$, we note that again a single temperature cannot fit all its lines. In particular, the line at 90.593\,GHz with $E_{\rm up} = 19$\,K gets overestimated regardless of the temperature assumed. This could be because of this line being (marginally) optically thick, therefore, we ignored this line and fixed the temperature to that of gGg$^{\prime}$-(CH$_2$OH)$_2$. A similar two-component temperature structure was found for glycolonitrile (HOCH$_2$CN) toward IRAS 16293 B (\citealt{Zeng2019}). Using this method the rest of the lines are explained reasonably well (i.e., within ${\sim}20-30\%$; see Fig. \ref{fig:model_spec}).

The column density was always treated as a free parameter. The uncertainty on column densities was measured from the same method explained in \cite{Nazari2022ALMAGAL} and the typical uncertainty from the fits is on the order of 20$\%$. Figure \ref{fig:model_spec} shows the final fitted model for each molecule. As this figure shows, there are still unidentified lines in the spectrum (see Sect. \ref{sec:detection} for more detail). In this process, the source velocity was fixed to $V_{\rm lsr} = 2.7$\,km\,s$^{-1}$ and a beam dilution of unity was assumed resulting in our column densities being representative of those within a beam. This is different from the assumption in PILS where the source size was set to $0.5\arcsec$. Nevertheless, this assumption does not change the conclusions as long as the lines are optically thin. This is because only the number of molecules and their ratios are of interest here. The number of molecules ($\mathcal{N}$) is constant regardless of the source size assumed. The number of molecules is equal to $N\times A$, where $A$ is the emitting area. Hence a decrease in the emitting area will increase the fitted column densities and vice versa such that the number of molecules stays the same as long as the lines stay optically thin (also see \citealt{vanGelder2022Deuteration}; \citealt{Nazari2023CGD}). However, we note that with this assumption we ignore any potential differences between the emitting areas of various molecules. This assumption can only be improved with higher angular resolution data than presented in this work.

\section{Results}
\label{sec:results}
\subsection{Deep search and the considered species}
\label{sec:detection}

\begin{figure*}
    \centering
    \includegraphics[width=18cm]{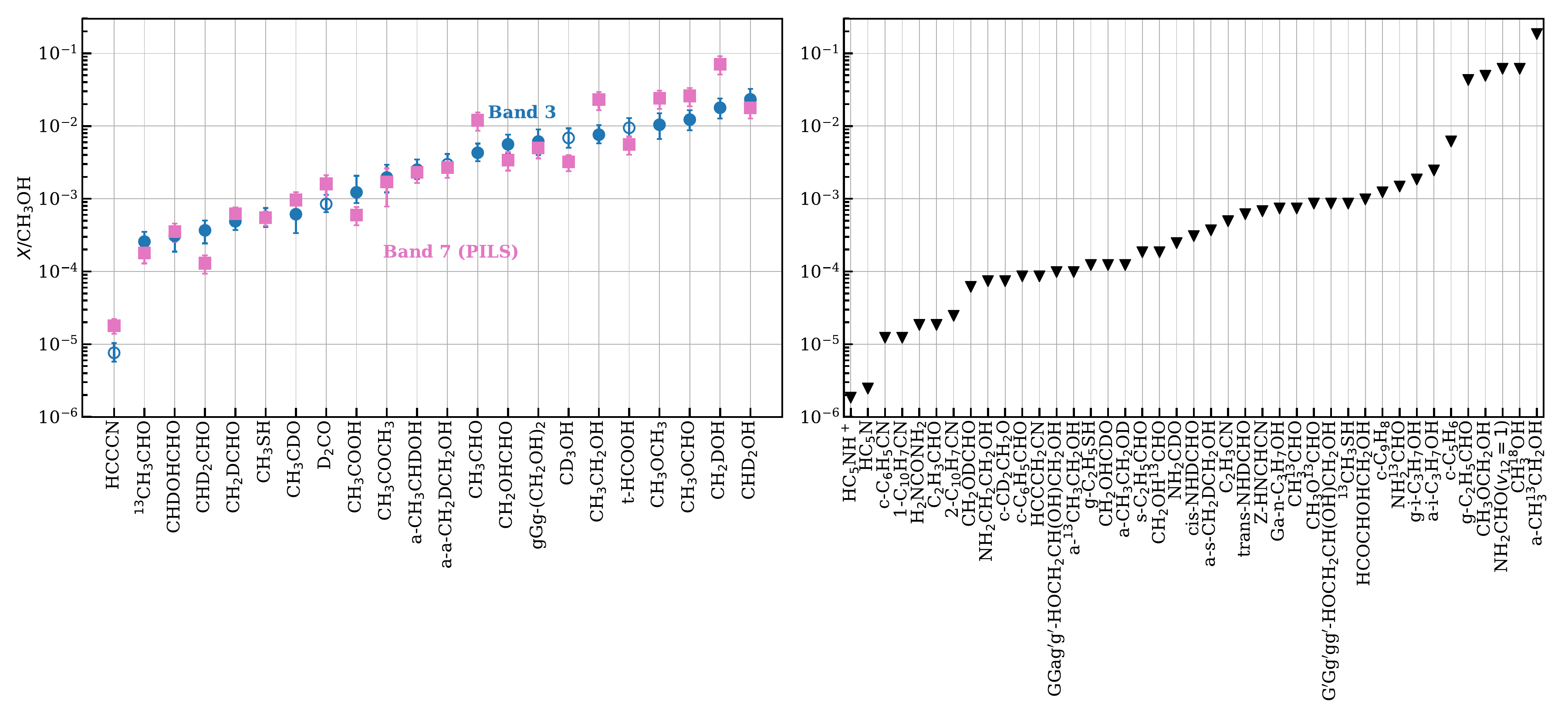}
    \caption{Left: column density ratios for the detected species with respect to methanol for our Band 3 data (blue). The same ratios from PILS in Band 7 (pink) are also shown for comparison (see the text for references). The hollow symbols show the species that only have one detected line. Right: ratios of the upper limits measured in this work (Table \ref{tab:non_detect}) with respect to methanol. In both panels, the species are ordered from left to right by increasing Band 3 ratios with respect to methanol.} 
    \label{fig:uppers}
\end{figure*}

The spectrum of IRAS16293 B at ${\sim}3$\,mm wavelengths is line-rich and crowded (see Fig. \ref{fig:overview} for an overview). In total 16 molecules are detected and 15 are tentatively detected (see Table \ref{tab:results}). Ratios of the detected molecules with respect to methanol are presented in left panel of Fig. \ref{fig:uppers} (see Sect. \ref{sec:ratios_PILS} for discussion on comparison with PILS). Among these species many `standard' complex organic molecules such as methanol (CH$_3$OH), ethanol (CH$_3$CH$_2$OH), ethyl cyanide (CH$_3$CH$_2$CN), acetaldehyde (CH$_3$CHO), methyl formate (CH$_3$OCHO), and dimethyl ether (CH$_3$OCH$_3$) with some of their isotopologues are apparent. The frequency range did not cover transitions of CH$_3$CN and HNCO or their isotopologues. We also detect one carbon chain molecule, cyanoacetylene (HCCCN). It should be noted that the frequency  range of the observations covers HC$_5$N transitions, but this carbon chain is not detected toward IRAS16293 B. It is, however, detected ${\sim}12\arcsec$ off-source to the west of IRAS16293 B (\citealt{Murillo2022}).

\begin{table}
\renewcommand{\arraystretch}{1.3}
    \caption{Fitted parameters for detected and tentatively detected Band 3 species toward IRAS 16293 B.}
    \label{tab:results}
    \resizebox{\columnwidth}{!}{\begin{tabular}{@{\extracolsep{1mm}}*{5}{l}}
          \toprule
          \toprule      
        Species & $N$ &  $T_{\rm ex} $ &  $N_{\rm{X}}/N_{\rm{CH}_{3}\rm{OH}}$ & \# Detected \\ 
        &$(\rm cm^{-2})$&(\rm K)&&lines \\
        \midrule     

$^{13}$CH$_3$OH & 1.2$^{+0.4}_{-0.2}$ $\times 10^{17}$ & 130$^{+30}_{-20}$ & 1.5$^{+0.7}_{-0.4}$ $\times 10^{-2}$ & 2 \\
CH$_2$DOH & 1.5$^{+0.1}_{-0.3}$ $\times 10^{17}$ & [170] & 1.8$^{+0.6}_{-0.5}$ $\times 10^{-2}$ & 3 \\
CHD$_2$OH & 1.9$^{+0.4}_{-0.3}$ $\times 10^{17}$ & 170$^{+40}_{-40}$ & 2.3$^{+0.9}_{-0.6}$ $\times 10^{-2}$ & 9 \\
CD$_3$OH & 5.6$^{+0.7}_{-0.9}$ $\times 10^{16}$ & [170] & 6.9$^{+2.4}_{-1.8}$ $\times 10^{-3}$ & 1 \\
CH$_3$CHO & 3.5$^{+0.2}_{-0.4}$ $\times 10^{16}$ & 70$^{+20}_{-10}$ & 4.3$^{+1.5}_{-1.0}$ $\times 10^{-3}$ & 3 \\
$^{13}$CH$_3$CHO & 2.1$^{+0.3}_{-0.3}$ $\times 10^{15}$ & [100] & 2.6$^{+0.9}_{-0.7}$ $\times 10^{-4}$ & 2 \\
CH$_2$DCHO & 4.0$^{+0.5}_{-0.5}$ $\times 10^{15}$ & [100] & 4.9$^{+1.7}_{-1.2}$ $\times 10^{-4}$ & 2 \\
CH$_3$CDO & 5.0$^{+2.0}_{-2.0}$ $\times 10^{15}$ & 110$^{+30}_{-30}$ & 6.1$^{+3.2}_{-2.8}$ $\times 10^{-4}$ & 3 \\
CHD$_2$CHO & 3.0$^{+0.5}_{-0.8}$ $\times 10^{15}$ & [100] & 3.7$^{+1.4}_{-1.2}$ $\times 10^{-4}$ & 3 \\
CH$_3$COOH & 1.0$^{+0.6}_{-0.2}$ $\times 10^{16}$ & 170$^{+30}_{-70}$ & 1.2$^{+0.8}_{-0.4}$ $\times 10^{-3}$ & 3 \\
CH$_2$OHCHO & 4.6$^{+0.6}_{-0.6}$ $\times 10^{16}$ & 280$^{+20}_{-40}$ & 5.6$^{+2.0}_{-1.4}$ $\times 10^{-3}$ & 6 \\
$^{13}$CH$_2$OHCHO & ${\sim}$2.0 $\times 10^{15}$ & [280] & ${\sim}$2.4 $\times 10^{-4}$ & 2 \\
CHDOHCHO & 2.5$^{+0.8}_{-0.8}$ $\times 10^{15}$ & 120$^{+50}_{-50}$ & 3.1$^{+1.5}_{-1.2}$ $\times 10^{-4}$ & 4 \\
CH$_3$CH$_2$OH & 6.2$^{+0.8}_{-0.7}$ $\times 10^{16}$ & 170$^{+20}_{-20}$ & 7.6$^{+2.7}_{-1.8}$ $\times 10^{-3}$ & 5 \\
a-a-CH$_2$DCH$_2$OH & 2.4$^{+0.5}_{-0.3}$ $\times 10^{16}$ & [170] & 3.0$^{+1.1}_{-0.7}$ $\times 10^{-3}$ & 1 \\
a-CH$_3$CHDOH & 2.1$^{+0.4}_{-0.3}$ $\times 10^{16}$ & [170] & 2.5$^{+1.0}_{-0.6}$ $\times 10^{-3}$ & 3 \\
CH$_3$OCH$_3$ & 8.5$^{+2.5}_{-2.5}$ $\times 10^{16}$ & 100$^{+20}_{-20}$ & 1.0$^{+0.5}_{-0.4}$ $\times 10^{-2}$ & 10 \\
CH$_3$OCHO & 1.0$^{+0.1}_{-0.2}$ $\times 10^{17}$ & 140$^{+20}_{-30}$ & 1.2$^{+0.4}_{-0.4}$ $\times 10^{-2}$ & 10 \\
aGg$^{\prime}$-(CH$_2$OH)$_2$ & ${\sim}$1.4 $\times 10^{17}$ & [160] & ${\sim}$1.7 $\times 10^{-2}$ & 15 \\
gGg$^{\prime}$-(CH$_2$OH)$_2$ & 5.0$^{+1.7}_{-1.4}$ $\times 10^{16}$ & 160$^{+40}_{-40}$ & 6.1$^{+2.9}_{-2.1}$ $\times 10^{-3}$ & 15 \\
D$_2$CO & 6.9$^{+0.6}_{-0.6}$ $\times 10^{15}$ & [100] & 8.5$^{+2.9}_{-1.9}$ $\times 10^{-4}$ & 1 \\
HCCCN & 6.2$^{+0.9}_{-0.8}$ $\times 10^{13}$ & [100] & 7.6$^{+2.8}_{-1.8}$ $\times 10^{-6}$ & 1 \\
CH$_3$CH$_2$CN & 1.4$^{+0.2}_{-0.1}$ $\times 10^{16}$ & [100] & 1.8$^{+0.7}_{-0.4}$ $\times 10^{-3}$ & 1 \\
NH$_2$CHO & 5.6$^{+1.2}_{-1.1}$ $\times 10^{16}$ & [300] & 6.8$^{+2.7}_{-1.9}$ $\times 10^{-3}$ & 1 \\
CH$_3$COCH$_3$ & 1.6$^{+0.6}_{-0.5}$ $\times 10^{16}$ & 130$^{+20}_{-20}$ & 2.0$^{+1.0}_{-0.7}$ $\times 10^{-3}$ & 17 \\
t-HCOOH & 7.7$^{+1.1}_{-1.1}$ $\times 10^{16}$ & [300] & 9.5$^{+3.4}_{-2.4}$ $\times 10^{-3}$ & 1 \\
c-C$_2$H$_4$O & 5.5$^{+0.7}_{-0.5}$ $\times 10^{15}$ & [100] & 6.7$^{+2.4}_{-1.5}$ $\times 10^{-4}$ & 2 \\
c-C$_2$H$_3$DO & 1.2$^{+0.2}_{-0.2}$ $\times 10^{15}$ & [100] & 1.5$^{+0.5}_{-0.4}$ $\times 10^{-4}$ & 2 \\
CH$_3$SH & 4.5$^{+0.5}_{-0.7}$ $\times 10^{15}$ & [100] & 5.5$^{+1.9}_{-1.4}$ $\times 10^{-4}$ & 2 \\
HOCH$_2$CN & 1.0$^{+0.1}_{-0.1}$ $\times 10^{15}$ & [100] & 1.2$^{+0.4}_{-0.3}$ $\times 10^{-4}$ & 2 \\

\bottomrule
        \end{tabular}}
        \tablefoot{Measured column densities toward the 0.5\arcsec offset position from B in a ${\sim} 1\arcsec$ beam. If a source size of 0.5$\arcsec$ was assumed, these column densities would increase by a factor of ${\sim}5$ (i.e., $\frac{0.5^2 + 1^2}{0.5^2}$). The major isotopologue of methanol is detected but its column density is calculated by scaling the $^{13}$CH$_3$OH column density by $^{12}$C/$^{13}$C $= 68$ (\citealt{Milam2005}). The FWHM for all molecules is ${\sim}1.8\pm 0.5$\,km~s$^{-1}$. Species whose excitation temperature is fixed have their temperature given in square brackets. The right-most column gives an estimate of the number of relatively unblended lines that are detected for each molecule. Those detected with only one line should be taken with caution. However, we note that all of these species are detected in PILS (see the text for references).}
\end{table}

This data set was specifically taken to search for large species due to the expected lower line density and line confusion in Band 3 in comparison with Band 7 used for PILS. Moreover, at the same excitation temperature heavier molecules have their Boltzmann distribution peak at lower frequencies which increases the chance of detecting them in Band 3. Particularly, the frequency windows were selected to search for glycerol (HOCH$_2$CH(OH)CH$_2$OH) and propanol (C$_3$H$_7$OH). Isopropanol (i-C$_3$H$_7$OH) and normal-propanol (n-C$_3$H$_7$OH) have been detected previously in the interstellar medium (\citealt{Belloche2022}; \citealt{Jimenez2022}), although not yet in IRAS16293 B (\citealt{Taquet2018}). In addition to these molecules, we searched for and detected large and complex species such as aGg’- and gGg’-ethylene glycol (CH$_2$OH)$_2$, glycolaldehyde (CH$_2$OHCHO), and acetone (CH$_3$COCH$_3$).  

\begin{figure*}
    \centering
    \includegraphics[width=19cm]{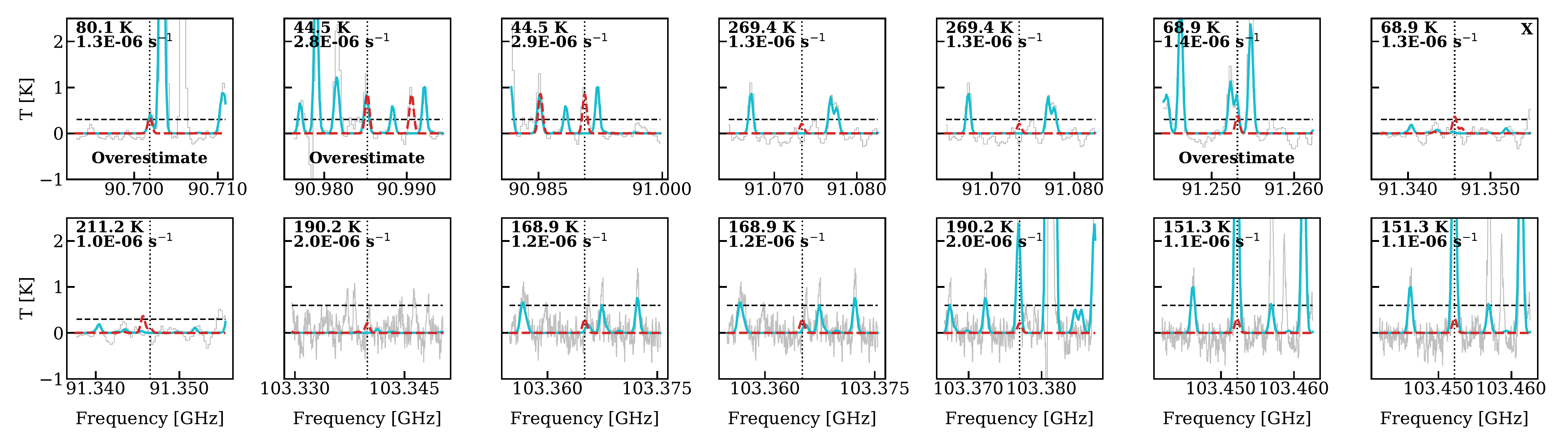}
    \caption{Lines of g-Isopropanol and the model for its upper limit in red dashed lines ($1.5\times10^{16}$\,cm$^{-2}$). Gray is the data and cyan is the total fitted model from the detected and tentatively detected species. The vertical dotted lines show the transition frequency of each line. The $E_{\rm up}$ and $A_{\rm ij}$ are printed on the top left of each panel. The line that is used to find the $3\sigma$ upper limit is indicated by an `X' on the top right. The horizontal dashed lines show the $3\sigma$ level. Only lines with $A_{\rm ij} > 10^{-6}$\,s$^{-1}$ and $E_{\rm up} < 300$\,K are shown.} 
    \label{fig:prop}
\end{figure*}

Table \ref{tab:non_detect} presents the upper limits of the molecules searched for, but not detected including glycerol and isopropanol. The largest molecule that we searched for is cyanonaphthalene (C$_{10}$H$_7$CN; \citealt{McNaughton2018}) and its upper limit is given in Table \ref{tab:non_detect}. Figures \ref{fig:prop}, \ref{fig:n_prop}, and \ref{fig:glyc} present the lines of g-Isopropanol, Ga-n-propanol, and G$^\prime$Gg$^\prime$gg$^\prime$-Glycerol with the model upper limits. We note that in Fig. \ref{fig:prop} the lines that seem to agree well with the data in the first two panels are well explained with other molecules (see the cyan line as the total fitted model for the detected or tentatively detected species) and thus, could not be the lines of g-Isopropanol. Moreover, the line at around 91.345\,GHz overestimates the data. The ratio of measured upper limits with respect to methanol are presented in right panel of Fig. \ref{fig:uppers}. The ratios span a range of ${\sim}5$ orders of magnitude from ${\sim}10^{-6}$ to ${\sim}0.1$ which is similar to the range seen for the detected molecules (left panel of Fig. \ref{fig:uppers}).

The line density of the lines detected at $\gtrsim 6\sigma$ level in Band 3 (${\sim}100$\,GHz) data is 1 per ${\sim}8.5$\,MHz (see Fig. \ref{fig:model_spec} for the variations in line density across the frequency range), while the line density is 1 per ${\sim}3.5$\,MHz in Band 7 (${\sim}345$\,GHz; \citealt{Jorgensen2016}). Therefore, as expected the spectrum has lower line density in Band 3 than Band 7, although only by a factor of ${\sim}2.5$, resulting in a yet relatively line-rich spectrum. Although many lines (${\sim}70$\%) in the Band 3 spectrum of IRAS16293 B are identified, we could not associate any simple or complex species to ${\sim}25-30\%$ of lines at the $\gtrsim 6\sigma$ level. Potentially more high resolution laboratory spectra are needed to identify those lines. This is particularly important for the (doubly) deuterated isotopologues of known COMs and larger COMs with more than 8 atoms.

\subsection{Fitting results for the detected species}

Derived column densities and excitation temperatures are given in Table \ref{tab:results}. The measured excitation temperatures span a range between ${\sim}50$ and ${\sim}300$\,K. Figure \ref{fig:Tex} presents the excitation temperatures where a determination was possible for molecules with sufficient detected lines. The species on the x-axis are roughly ordered by increasing binding energies in the ice from left to right (\citealt{Minissale2022}; \citealt{Ligterink2023}). The error bars are too large to robustly confirm whether there is any correlation between the excitation temperature and binding energy.

The column densities measured here span a range of ${\sim}4$ orders of magnitude. The most abundant molecule is methanol, after that (and its isotopologues), CH$_3$OCHO and CH$_3$OCH$_3$ are found to be the second and third most abundant species as found in many other protostellar systems (\citealt{Coletta2020}; \citealt{Chen2023}). The column density of CH$_3$OH is determined by scaling the $^{13}$CH$_3$OH column density by $^{12}$C/$^{13}$C of 68 (\citealt{Milam2005}) and is found to be $8.2 \times 10^{18}$\,cm$^{-2}$ in the ${\sim}1\arcsec$ beam. We note that the $^{13}$CH$_3$OH lines used for measurement of column density of this molecule are relatively weak and hence optically thin. They either have a large $E_{\rm up}$(${\sim}500$\,K) or a low $A_{\rm ij}$ (${\sim}6\times10^{-8}$\,s$^{-1}$). Therefore, $^{13}$CH$_3$OH in this study should robustly determine the column density of the major isotopologue without the need for CH$_3^{18}$OH. The least abundant molecule at the 0.5$\arcsec$ offset location is HCCCN with a column density of $6.2 \times 10^{13}$\,cm$^{-2}$ (see \citealt{Murillo2022} for its map).

\subsection{Glycolonitrile and ethylene oxide}

Here we focus on the tentative detection of HOCH$_2$CN (glycolonitrile) and c-C$_2$H$_4$O (ethylene oxide). These two molecules are among the less common species studied toward protostars in Table \ref{tab:results}. Glycolonitrile is an interesting interstellar molecule to study given that it is a rarely observed prebiotic molecule. Ethylene oxide is an interesting molecule because it is the only species in Table \ref{tab:results} with a cyclic structure.

Glycolonitrile was detected toward IRAS 16293 B using lower frequency ($\lesssim 266$\,GHz) observations than PILS (\citealt{Zeng2019}). Later it was also detected in PILS by \cite{Ligterink2021} mainly toward the half-beam offset position (${\sim}0.25\arcsec$ offset from the continuum peak) and not the full-beam offset position ($0.5\arcsec$ offset). It is interesting that in the Band 3 data HOCH$_2$CN is tentatively detected toward the full-beam offset position of PILS. This could be due to the larger beam size of the Band 3 observations and inclusion of the hotter gas close to the protostar in the beam given that the binding energy of this molecule is relatively high (${\sim}10\,400$\,K; \citealt{Ligterink2023}).

Our tentative column density ratio of HOCH$_2$CN/CH$_3$OH (${\sim}10^{-4}$) agrees well with what \cite{Ligterink2021} found for this source in Band 7. Moreover, our column density for HOCH$_2$CN, after correction for beam dilution (see Sect. \ref{sec:col}), is within a factor of ${\sim}2$ of what \cite{Zeng2019} find for their warm component of the same source from their Band 3 data. Moreover, glycolonitrile has been (tentatively) detected toward other objects such as the Serpens SMM1-a protostar and the G+0.693-0.027 molecular cloud with similar HOCH$_2$CN/CH$_3$OH ratios of a few $10^{-4}$ (\citealt{Requena2006}; \citealt{Ligterink2021}; \citealt{Rivilla2022}). A recent study searched for its minor isotopologues in IRAS16293B and SMM1-a but resulted in non-detections (\citealt{Margules2023}).  

Ethylene oxide has been detected toward several objects mainly high-mass protostars but also pre-stellar cores and the comet 67P (\citealt{Dickens1997}; \citealt{Nummelin1998}; \citealt{Ikeda2001}; \citealt{Requena2008}; \citealt{Bacmann2019}; \citealt{Drozdovskaya2019}). This molecule has also been detected toward IRAS16293 A and B by the PILS (\citealt{Lykke2017}; \citealt{Manigand2020}). The ratio for ethylene oxide to methanol from this work is ${\sim}7\times 10^{-4}$ which agrees well with the ratio from PILS of ${\sim}3-6 \times 10^{-4}$ (\citealt{Lykke2017}; \citealt{Jorgensen2016, Jorgensen2018}). Its deuterated species were studied and detected toward IRAS16293 B by \cite{Muller2023_2, Muller2023}. We also tentatively detect one of its deuterated species, c-C$_2$H$_3$DO, in our Band 3 data. The ratio of this molecule with respect to methanol in our data is ${\sim}1-2\times 10^{-4}$ which agrees well with the same ratio in PILS (${\sim}9 \times 10^{-5}$; \citealt{Jorgensen2018}; \citealt{Muller2023_2}).

\section{Discussion}
\label{sec:discussion}
\subsection{Dust optical depth} 
\label{sec:tau}
In this section we calculate the continuum optical depth at ${\sim}348.815$\,GHz (corresponding to ALMA Band 7) and at ${\sim}91$\,GHz (corresponding to ALMA Band 3). The continuum optical depth as a zeroth-order approximation is given by (see \citealt{Rivilla2017} and \citealt{vanGelder2022})

\begin{equation}
    \tau_{\nu} = -\ln\left(1- \frac{F_{\nu}}{\Omega_{\rm beam}B_{\nu}(T_{\rm dust})}\right),
    \label{eq:dust_tau}
\end{equation}

\noindent where $F_{\nu}$ is the continuum flux density within the beam, $\Omega_{\rm beam} = \pi \theta_{\rm min} \theta_{\rm maj}/(4\ln(2))$ is the beam solid angle with $\theta_{\rm min}$ and $\theta_{\rm maj}$ as the beam minor and major axes, $B_{\nu}$ is the Planck function and $T_{\rm dust}$ is the dust temperature. We took the continuum image of the PILS survey at $\nu {\sim}348.815$\,GHz from the ALMA archive and found $F_{\nu}$ within the beam of those observations at the peak of the continuum and at the ${\sim}0.5\arcsec$ offset position where the spectrum was extracted. For this measurement we used CASA (\citealt{McMullin2007}) version 6.5.2.26 and found the continuum flux density as ${\sim}1.16$\,Jy\,beam$^{-1}$ and ${\sim}0.55$\,Jy\,beam$^{-1}$ at the peak and the offset position. We did the same for the Band 3 data at a frequency of ${\sim}91$\,GHz and found the Band 3 continuum flux density as ${\sim}9.07 \times 10^{-2}$\,Jy\,beam$^{-1}$ and ${\sim}5.25 \times 10^{-2}$\,Jy\,beam$^{-1}$ at the peak and the offset positions, respectively. Next, we calculated $\tau_{\rm dust}$ for a range of dust temperatures that are feasible for IRAS16293B as suggested by \cite{Jacobsen2018}. 

\begin{figure}
  \resizebox{\columnwidth}{!}{\includegraphics{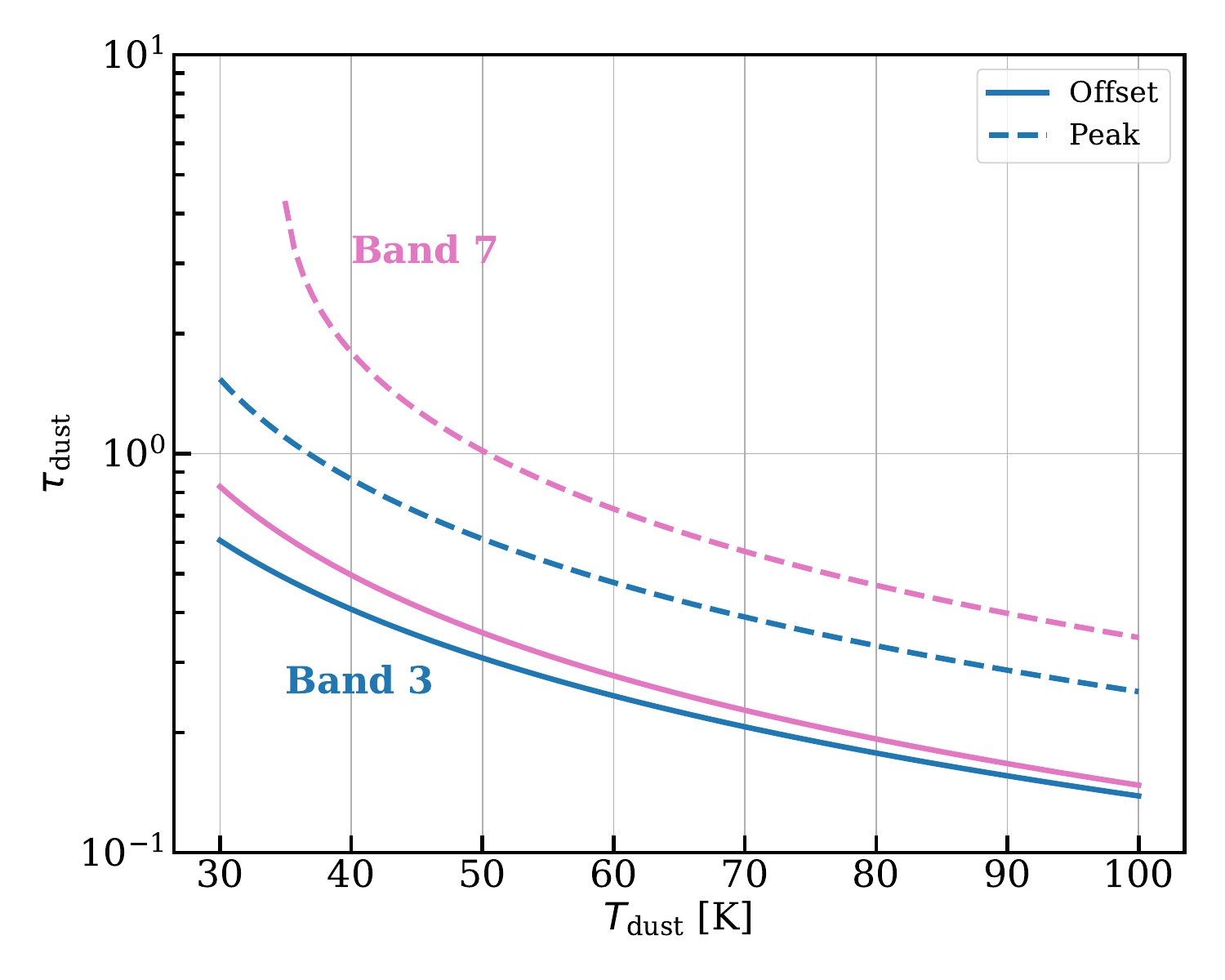}}
  \caption{Continuum optical depth as a function of temperature at the peak of the continuum (dashed) and the 0.5$\arcsec$ offset position where the spectrum is extracted (solid line). Blue shows the optical depth for the Band 3 continuum at frequency of ${\sim}91$\,GHz and pink shows the PILS continuum at frequency of ${\sim}348.815$\,GHz.}
  \label{fig:tau}
\end{figure}

Figure \ref{fig:tau} presents the continuum optical depth for the various temperatures. The temperature of the inner regions based on models of \cite{Jacobsen2018} is ${\sim} 90$\,K. At 90\,K both the PILS and Band 3 peak continuum are marginally optically thick with the PILS continuum having an optical depth of ${\sim}1.5$ times larger. However, at the location off-source the dust optical depth for the PILS continuum is almost the same as that of Band 3 which is ${\sim}0.15$. Therefore, it is safe to assume that the dust at the offset location at $T_{\rm dust}=90$\,K is almost completely optically thin in Band 3 and Band 7 data sets. Even taking the unrealistic worst case scenario of $T_{\rm dust}=30$\,K in Fig. \ref{fig:tau} and assuming that all the dust is in a column between the observer and the protostar, the difference between Band 3 and Band 7 dust attenuation (i.e., $e^{-\tau}$) is at most around 25\% at the offset location. 


\subsection{Comparison with PILS}
In this section we compare the excitation temperatures and column densities of the same molecules detected in the PILS Band 7 survey and this work.

\subsubsection{Excitation temperature}
\label{sec:onion}

The measured excitation temperatures are presented in Fig. \ref{fig:Tex}. The species are roughly ordered by increasing binding energy (a measure of how strongly bound a molecule is in ices) from left to right. However, given that the uncertainties are relatively large, the number of detected lines for each species is small, and the lines are biased toward lower $E_{\rm up}$, it is not possible to identify a clear trend between the binding energies (or sublimation temperatures) and excitation temperatures. Another effect that can complicate the picture and change Fig. \ref{fig:Tex} is that the molecules with lower binding energies than water are expected to desorb with water if initially mixed with it (\citealt{Collings2004}; \citealt{Busch2022}; \citealt{Garrod2022}). This would be inline with the measured excitation temperatures of the molecules toward the left hand side of Fig. \ref{fig:Tex} being at around ${\sim}100$\,K (i.e., the desorption temperature of water). Therefore, a strong trend might not be expected in Fig. \ref{fig:Tex}.

Nevertheless, our excitation temperatures mostly agree with those found by \cite{Jorgensen2018} for IRAS16293 B. The only exceptions are CH$_3$OCHO and CH$_3$CH$_2$OH where their excitation temperatures are found to be higher in PILS. This could be due to the lines that are covered in the Band 3 data having lower $E_{\rm up}$ than those covered in the Band 7 data. For example, if the two lines of CH$_3$CH$_2$OH with $E_{\rm up}{\sim}80-90$\,K are ignored, a fit at $T_{\rm ex} {\sim}240$\,K (i.e., agreeing with PILS temperature) can match the brightness temperatures of the rest of the lines within ${\sim}40-50\%$. Our excitation temperatures also agree well with what is found for the companion source, IRAS16293A, in PILS (\citealt{Manigand2020}).

\subsubsection{Column density}
\label{sec:col}

Direct comparison of the column densities from the PILS and this work should be made with caution due to the different beam sizes in the two sets of observations. Therefore, before comparison, column densities of Band 7 are corrected to match the Band 3 results. PILS studies use a source size of 0.5$\arcsec$ in their analysis while we find the column densities averaged over the Band 3 beam. Our assumption corresponds to the Band 3 emission filling the beam uniformly (i.e., $\theta_{\rm s}>>\theta_{\rm b}$). The PILS column densities are converted to an average over the PILS beam (or filling the PILS beam uniformly) by multiplication with their beam dilution factor of $\frac{0.5^2}{0.5^2 + 0.5^2} = 0.5$. After this modification, we also investigated how the different beam sizes in PILS and our study can affect the comparison (assuming that the column densities are averaged over the respective beams). Figure \ref{fig:beams} presents a two dimensional Gaussian distribution with two circular regions representing the beams of Band 3 and Band 7. Assuming a uniform beam, we calculated the mean in the two beams and found around a 10\% difference between these two, which is smaller than the typical uncertainties on the column densities and thus is ignored.             

\begin{figure}
  \resizebox{\columnwidth}{!}{\includegraphics{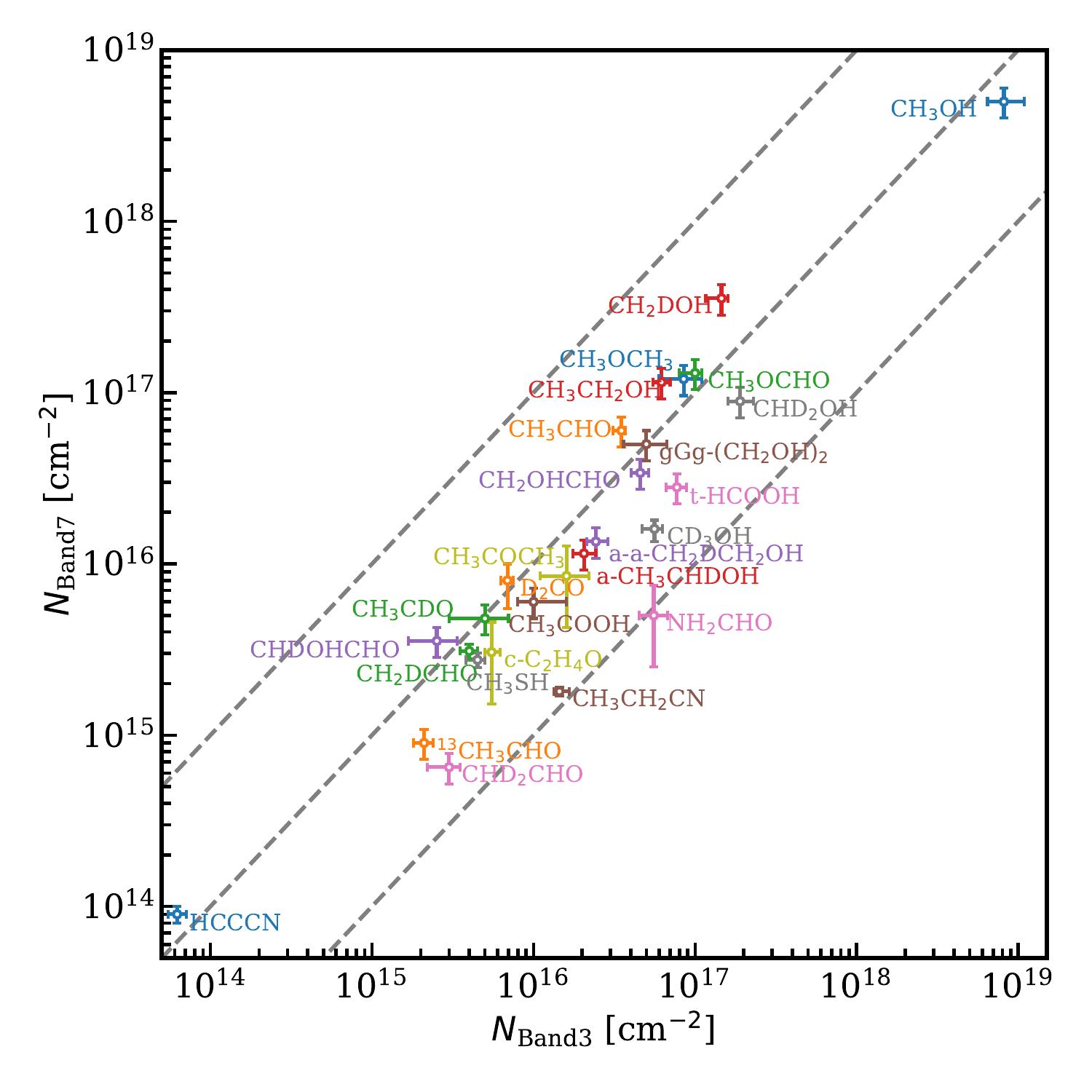}}
  \caption{Column density of the various molecules from the ALMA Band 7 observations (PILS; ${\sim}345$\,GHz) as a function of those from Band 3 (${\sim}100$\,GHz). Beam dilution is corrected for the column densities in this figure. Dashed lines present where the values of y-axis are the same as values of x-axis, where they are 10 times higher and 10 times lower than the values of x-axis.}
  \label{fig:cc}
\end{figure}

Figure \ref{fig:cc} presents the column densities from PILS and this work after correction for beam dilution. The column densities of PILS are taken from \cite{Coutens2016} (NH$_2$CHO), \cite{Jorgensen2016} (CH$_2$OHCHO, CHDOHCHO, gGg-(CH$_2$OH)$_2$, CH$_3$COOH), \cite{Calcutt2018} (CH$_3$CH$_2$CN, HC$_3$N), \cite{Jorgensen2018} (CH$_3$OH, $^{13}$CH$_3$CHO, CH$_3$CDO, CH$_3$CHO, CH$_2$DOH, CH$_3$CH$_2$OH, a-CH$_3$CHDOH, CH$_3$OCH$_3$, CH$_3$OCHO, a-a-CH$_2$DCH$_2$OH, and t-HCOOH), \cite{Persson2018} (D$_2$CO), \cite{Manigand2020} (CH$_2$DCHO), \cite{Drozdovskaya2022} (CHD$_2$OH), \cite{Ilyushin2022} (CD$_3$OH), \cite{Drozdovskaya2018} (CH$_3$SH), \cite{Ferrer2023} (CHD$_2$CHO), and \cite{Lykke2017} (CH$_3$COCH$_3$). In addition, we do not include $^{13}$CH$_2$OHCHO and gGg-(CH$_2$OH)$_2$ in Fig. \ref{fig:cc} because our measured column densities are approximate. Moreover, the column density of CH$_3$COOH in PILS was found from old spectroscopic data (\citealt{Jorgensen2016}). To avoid bias we re-did the fit to CH$_3$COOH in the Band 7 data using new spectroscopic data (\citealt{Ilyushin2013}; see Appendix \ref{sec:spec_data}), and found that the column density in \cite{Jorgensen2016} was underestimated by a factor of ${\sim}2$. Therefore, in this work we refer to the updated column density of CH$_3$COOH using the new spectroscopic data (${\sim}1.2\times 10^{16}$).

Figure \ref{fig:cc} shows that the column densities from the Band 3 observations generally agree with those of PILS. There are a few data points where the Band 3 column densities are ${\gtrsim}3$ times higher than those of Band 7. Among those molecules NH$_2$CHO and CH$_3$CH$_2$CN seem to be outliers which have Band 3 column densities that are 10 times higher than Band 7. However, because the column densities of these two molecules in this work are measured based on only one line those values should be taken with caution. Therefore, we exclude these two molecules from further analysis. 

Taking only the molecules with at least 3 detected lines in Band 3 (see Table \ref{tab:results}), the average of ($\log_{10}$ of) the Band 3 column densities weighted by the uncertainty on each data point is $4.4 \times 10^{16}$\,cm$^{-2}$. This value for Band 7 column densities (measured using many more lines per molecule due to the larger frequency coverage) is $2.7 \times 10^{16}$\,cm$^{-2}$. These two agree well within the uncertainties. Moreover, the scatter around the line of $N_{\rm Band 3} = N_{\rm Band 7}$ is a factor of 1.9 below the line and 1.8 above the line. Therefore, it can be concluded that there is a tight, one-to-one correlation between the two column densities with a scatter of less than a factor of 2. This agrees well with the low dust optical depth found for the Band 3 and Band 7 observations at the offset location where the spectra were extracted.

\subsubsection{Ratios}
\label{sec:ratios_PILS}

\begin{figure}
    \resizebox{\columnwidth}{!}{\includegraphics{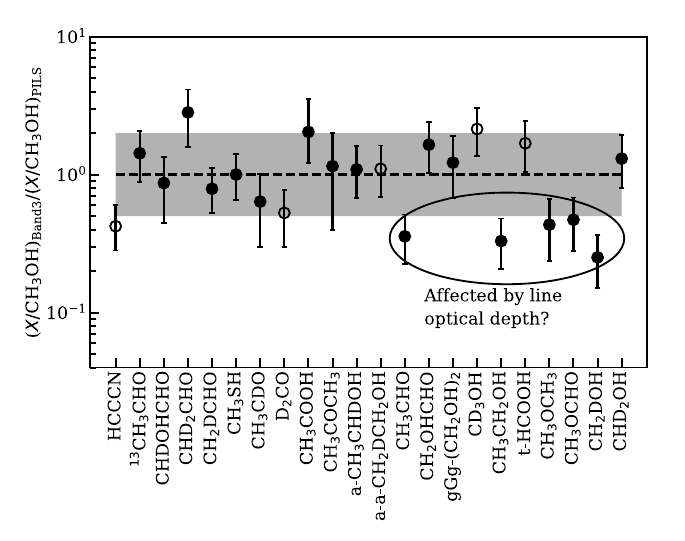}}
    \caption{Ratio between Band 3 and PILS column density ratios with respect to methanol (i.e., the ratio of blue to pink from the left panel of Fig. \ref{fig:uppers}). The horizontal dashed line indicates where the ratio between Band 3 and PILS are the same. The shaded gray area indicates the region with a factor of 2 difference between Band 3 and PILS results. The species are ordered in the same way as left panel of Fig. \ref{fig:uppers}. The hollow symbols show the species that have only one detected line.} 
    \label{fig:ratios}
\end{figure}

The column density ratios are normally more informative than absolute column densities because the latter depends on the beam and the assumed beam dilution factor. Left panel of Fig. \ref{fig:uppers} shows that the ratios with respect to methanol from Band 3 agree well with those from Band 7 (PILS) for almost all molecules. Figure \ref{fig:ratios} presents the ratio between Band 3 and Band 7 results for individual molecules. This figure shows even more clearly that the results from the Band 3 data (this work) generally agree within a factor of ${\sim}2$ with the results from the Band 7 data. Four molecules show larger than a factor of 2 difference between the two data sets and are marked with an ellipse. This could be due to the lines of these species becoming optically thick. This argument is more convincing for CH$_3$CHO and CH$_3$CH$_2$OH given that there is a good agreement between Band 3 and Band 7 results for their minor isotopologues. Moreover, the value for CH$_2$DOH is likely uncertain given that no single temperature could be fitted to its lines and hence the assumed $T_{\rm ex}$ does not fit all its lines equally well in this work (see Sect. \ref{sec:spec_modeling}). The slight variations between the Band 3 and Band 7 ratios could be due to the smaller number of lines covered in the Band 3 data set compared with the PILS data set, which also sets a looser constraints on the excitation temperatures. Another factor in these variations could be the different beam sizes in the two data sets (see Sect. \ref{sec:col}).

We find a good match between Band 3 and Band 7 data when comparing the ratios of the various COMs relative to methanol. The molecules that have column densities with a factor of $>2$ different between Band 3 and Band 7 in Fig. \ref{fig:cc} (i.e., t-HCOOH, CD$_3$OH, and CHD$_2$CHO), show a good agreement (factor of $< 2$) when their column density ratios with respect to methanol are compared between the two sets of observations.     

\subsection{Comparison with other studies}

In this section we put some of the measured ratios (Fig. \ref{fig:uppers}) in context of other works. We mainly focus on the upper limit ratios because similar comparisons have been made in the PILS papers between the measured column density ratios of IRAS 16293 B and other sources. The ratios of aGg$^{\prime}$- and gGg$^{\prime}$-(CH$_2$OH)$_2$ to CH$_3$OH from our observations is between ${\sim}4\times 10^{-3}$ and ${\sim}2 \times 10^{-2}$. The laboratory experiments and Monte Carlo simulations found (CH$_2$OH)$_2$/CH$_3$OH ratio to be ${\sim}10^{-2}-9\times10^{-2}$ and ${\sim}2 \times 10^{-2}-4\times 10^{-2}$ (\citealt{Fedoseev2015}) which agrees with our observations. Moreover, \cite{Fedoseev2017} found a ratio of glycerol/ethylene glycol upper limit of around 0.01 in laboratory experiments which from the ethylene glycol/methanol ratio of \cite{Fedoseev2015} gives glycerol/methanol upper limit of between $10^{-4}$ and $9\times10^{-4}$. Our upper limit ratios of GGag$^\prime$g$^\prime$- and G$^\prime$Gg$^\prime$gg$^\prime$-glycerol to methanol are ${<}10^{-4}$ and ${<}8 \times 10^{-4}$, which are on the same order of magnitude as reported in laboratory experiments and Monte Carlo simulations (\citealt{Fedoseev2015, Fedoseev2017}).

Propanal was found to form n-propanol (\citealt{Qasim2019,Qasim2019_Cocktail}), and both are not detected in the Band 3 data. However, we note that propanal is detected toward IRAS 16293 B by PILS (\citealt{Lykke2017}) and \cite{Qasim2019} found that the ratio of their upper limit ratio for n-propanol/propanal for IRAS 16293 B is consistent with their laboratory experiments. Our upper limit ratio of Ga-n-propanol/methanol is ${<}7\times 10^{-4}$ which is a factor of ${\sim}6$ lower than the detected ratio toward G+0.693-0.027 (${\sim}4\times 10^{-3}$; \citealt{Jimenez2022} and methanol from \citealt{Rodriguez2021}). Our upper limit ratios of g-, a-Isopropanol, and Ga-n-propanol with respect to methanol agree within a factor of about 2 with those of detected toward Sgr B2(N2b) (${\sim}0.002$; \citealt{Belloche2022}). Band 3 ratios of s- and g-propanal/methanol are ${<}2 \times 10^{-4}$ and ${<}0.04$, respectively. These are in agreement with the abundance ratios of propanal/methanol in TMC-1 (around 0.01, \citealt{Agundez2023}) and PILS (around $2\times 10^{-4}$; \citealt{Lykke2017}). Moreover, the abundance of indene with respect to H$_2$ toward TMC-1 (\citealt{Cernicharo2021_indene}), is on the same order of magnitude as our upper limit ratio of indene/H$_2$, assuming a lower limit on H$_2$ column density of $10^{25}$\,cm$^{-2}$ from \cite{Jorgensen2016} for IRAS 16293 B. However, using the same lower limit column density for H$_2$ toward IRAS 16293 B, our upper limit abundances of 1-CNN, 2-CNN and benzonitrile are a factor of ${\sim}2-10$ lower than what is found for TMC-1  (\citealt{Gratier2016}; \citealt{McGuire2021}).

The Band 3 upper limit ratio of urea/methanol is ${<}2\times 10^{-5}$ which is inline and on the lower end of the observed range (either upper limit or detected) in the literature toward SgrB2 (\citealt{Belloche2019}), NGC 6334I (\citealt{Ligterink2020}), and G+0.693-0.027 systems (\citealt{Zeng2023}). The upper limit NH$^{13}$CHO/CH$_3$OH in this work is consistent with the detected ratios toward NGC 6334I (\citealt{Ligterink2020}; methanol from \citealt{Bogelund2018}) and G+0.693-0.027 systems (\citealt{Zeng2023}; methanol from \citealt{Rodriguez2021}). Moreover, the upper limit ratios of z-cyanomethanimine and glyceraldehyde to methanol toward G+0.693-0.027 (\citealt{Jimenez2020}) are consistent with our upper limit ratios. However, the ratio of ethanolamine/methanol toward this source (\citealt{Rivilla2021}) was found to be around one order of magnitude higher than Band 3 upper limit measurement of IRAS 16293 B. The discussed ratios in this section may be generally higher in G+0.693-0.027 than IRAS 16293 B. Given the small sample size and the upper limit nature of our results, it is not possible to draw further conclusions on the significance of these differences and similarities. In particular, given that the upper limits reported here depend on the assumed excitation temperature and more importantly the lines covered in our data.

\section{Conclusions}
\label{sec:conclusions}

We analyzed the deep ALMA Band 3 (${\sim}100$\,GHz) data of IRAS16293 B in this work. We searched for large organic species in this data set and derived the corresponding column densities and excitation temperatures of various oxygen-, nitrogen-, and sulfur-bearing molecules. Below are the main conclusions of this work.

\begin{itemize}
    \item The line density for lines detected at a $\gtrsim 6\sigma$ level in Band 3 observations is 1 per ${\sim}8.5$\,MHz which is only ${\sim}2.5$ times lower than that of PILS: the spectrum is relatively rich and crowded even at ${\sim}3$\,mm observations.
    \item We detect around 31 molecules (including minor isotopologues), thereof ${\sim}15$ tentatively, in the Band 3 data set. These include O-bearing COMs such as CH$_3$OH, CH$_2$OHCHO, CH$_3$OCH$_3$, CH$_3$OCHO, gGg-(CH$_2$OH)$_2$, CH$_3$COCH$_3$, and c-C$_2$H$_4$O. We also tentatively detect a few N- and S-bearing species such as HOCH$_2$CN and CH$_3$SH. 
    \item We search for many large COMs among which are glycerol and isopropanol, but we do not detect them. The upper limits on the 41 non-detected species are also provided, which generally agree with the previous laboratory experiments and observations.
    \item In the Band 3 spectrum, ${\sim}25-30\%$ of all lines at $\gtrsim 6\sigma$ level are not identified. This points to the need for additional spectroscopic information. 
    \item We find a good agreement between column densities of Band 3 and Band 7 observations with a scatter of less than a factor of 2. Moreover, the Band 3 ratios with respect to methanol agree within a factor of ${\sim}2$ with those from the Band 7 observations.
    \item We conclude that around IRAS16293 B dust optical depth does not affect the column densities and the ratios of various molecules, especially for the spectrum extracted from a position off source (where the dust column density is lower than on-source). 
\end{itemize}

\begin{acknowledgements}
    We thank the referee for the helpful and constructive comments. We thank A. Hacar for his invaluable help with the data reduction. Astrochemistry in Leiden is supported by the Netherlands Research School for Astronomy (NOVA), by funding from the European Research Council (ERC) under the European Union’s Horizon 2020 research and innovation programme (grant agreement No. 101019751 MOLDISK), and by the Dutch Research Council (NWO) grant 618.000.001. Support by the Danish National Research Foundation through the Center of Excellence “InterCat” (Grant agreement no.: DNRF150) is also acknowledged. J.K.J. acknowledges support from the Independent Research Fund Denmark (grant number 0135-00123B). M.N.D. acknowledges the Swiss National Science Foundation (SNSF) Ambizione grant number 180079, the Center for Space and Habitability (CSH) Fellowship, and the IAU Gruber Foundation Fellowship. G.F. acknowledges the financial support from the European Union’s Horizon 2020 research and innovation program under the Marie Sklodowska-Curie grant agreement No. 664931. R.T.G. acknowledges funding from the National Science Foundation Astronomy \& Astrophysics program (grant number 2206516). H.S.P.M. acknowledges support by the Deutsche Forschungsgemeinschaft (DFG) via the collaborative research grant SFB 956 (project ID 184018867). S.F.W. acknowledges the financial support of the SNSF Eccellenza Professorial Fellowship (PCEFP2\textunderscore181150). This paper makes use of the following ALMA data: ADS/JAO.ALMA\#2017.1.00518.S and \#2013.1.00278.S. ALMA is a partnership of ESO (representing its member states), NSF (USA) and NINS (Japan), together with NRC (Canada), MOST and ASIAA (Taiwan), and KASI (Republic of Korea), in cooperation with the Republic of Chile. The Joint ALMA Observatory is operated by ESO, AUI/NRAO and NAOJ. The National Radio Astronomy Observatory is a facility of the National Science Foundation operated under cooperative agreement by Associated Universities, Inc.   
\end{acknowledgements}

%
%

\bibliographystyle{aa}
\bibliography{IRASB_band3}

\begin{appendix}

\section{Spectroscopic data}
\label{sec:spec_data}

The spectroscopic information are summarized in Table \ref{tab:spec_info}. The vibrational correction factor is assumed as 1 for all molecules except for the following species. Vibrational correction factor of 1.457 is used for CH$_2$DOH at a temperature of 300\,K (\citealt{Lauvergnat2009}; \citealt{Jorgensen2018}). For CH$_2$OHCHO, a factor of 2.86 is adopted at 300K to take into account the higher (than ground state) vibrational states which are calculated in the harmonic approximation. For $^{13}$CH$_2$OHCHO and CHDOHCHO a vibrational factor of 2.8 is used at a temperature of 300\,K (\citealt{Jorgensen2016}). An upper limit vibrational factor of 2.824 at 300\,K (\citealt{Durig1975}; \citealt{Jorgensen2018}) is assumed for a-a-CH$_2$DCH$_2$OH and a-CH$_3$CHDOH. Moreover, following \cite{Jorgensen2018}, we multiply the column densities of these two species by an additional factor of ${\sim}2.69$ to account for the presence of the gauche conformer. We assume a vibration correction factor of 4.02 for aGg-(CH$_2$OH)$_2$ and gGg-(CH$_2$OH)$_2$ and add an additional factor of ${\sim}1.38$ to include the contribution from the higher gGg conformer at 300\,K (\citealt{Muller2004}; \citealt{Jorgensen2016}). A vibrational correction factor of 1.09 at 100\,K is used for HCCCN (\citealt{Mallinson1976}; \citealt{Calcutt2018}). For CH$_3$CH$_2$CN a vibrational correction factor of 1.113 is assumed at 100\,K. A vibrational correction factor of 1.5 is assumed for NH$_2$CHO at 300\,K. For t-HCOOH a vibrational correction factor of 1.103 at 300\,K is assumed (\citealt{Perrin2002}; \citealt{Baskakov2006}; \citealt{Jorgensen2018}).

\begin{table}
\renewcommand{\arraystretch}{1.3}
    \caption{Spectroscopic information}
    \label{tab:spec_info}
    \resizebox{\columnwidth}{!}{\begin{tabular}{@{\extracolsep{1mm}}*{4}{l}}
          \toprule
          \toprule      
        Name & Species & Catalog & References \\
        \midrule     

Methanol & $^{13}$CH$_3$OH & CDMS & \cite{Xu1997} \\
Methanol &CH$_2$DOH & JPL & \cite{Pearson2012}\\
Methanol &CHD$_2$OH & CDMS & \cite{Drozdovskaya2022} \\
Methanol &CD$_3$OH & CDMS & \cite{Ilyushin2022}\\
Acetaldehyde & CH$_3$CHO & JPL & \cite{Kleiner1996}\\
Acetaldehyde &$^{13}$CH$_3$CHO & CDMS & \cite{Margules2015}\\ 
Acetaldehyde &CH$_2$DCHO & CDMS & \cite{Coudert2019} \\
Acetaldehyde &CH$_3$CDO &  CDMS & \cite{Coudert2019}\\
Acetaldehyde &CHD$_2$CHO & JPL format & \cite{Ferrer2023}\\
Acetic acid & CH$_3$COOH & CDMS & \cite{Ilyushin2013} \\
Glycolaldehyde & CH$_2$OHCHO & CDMS &   M\"{u}ller 2021, unpublished\\
Glycolaldehyde &$^{13}$CH$_2$OHCHO & CDMS & \cite{Haykal2013}\\ 
Glycolaldehyde &CHDOHCHO & CDMS & \cite{Bouchez2012} \\
Ethanol & CH$_3$CH$_2$OH & CDMS & \cite{Pearson2008}; \cite{Muller2016} \\
Ethanol &a-a-CH$_2$DCH$_2$OH & CDMS & \cite{Walters2015} \\
Ethanol &a-CH$_3$CHDOH & CDMS & \cite{Walters2015} \\
Dimethyl ether & CH$_3$OCH$_3$ & CDMS & \cite{Endres2009}  \\     
Methyl formate & CH$_3$OCHO & JPL & \cite{Ilyushin2009} \\
aGg$^{\prime}$-ethylene glycol& aGg$^{\prime}$-(CH$_2$OH)$_2$ & CDMS & \cite{Christen1995}; \cite{Christen2003} \\
gGg$^{\prime}$-ethylene glycol & gGg$^{\prime}$-(CH$_2$OH)$_2$ & CDMS & \cite{Christen2001}; \cite{Muller2004}\\
Formaldehyde & D$_2$CO & CDMS & \cite{Dangoisse1978}; \cite{Bocquet1999}\\
Cyanoacetylene & HCCCN & CDMS & \cite{Zafra1971}; \cite{Creswell1977}; \\
&&&\cite{Yamada1995}; \cite{Thorwirth2000} \\
Ethyl cyanide & CH$_3$CH$_2$CN & CDMS & \citealt{Pearson1994}; \citealt{Fukuyama1996};\\
&&&\citealt{Brauer2009}\\ 
Formamide & NH$_2$CHO & CDMS & \citealt{Kukolich1971}; \citealt{Hirota1974}; \\
&&&\citealt{Kryvda2009}; \citealt{Motiyenko2012} \\
Acetone & CH$_3$COCH$_3$ & JPL & \cite{Groner2002}; \cite{Ordu2019} \\
Formic acid &t-HCOOH & CDMS & \cite{Winnewisser2002}\\
Ethylene oxide & c-C$_2$H$_4$O & CDMS & \cite{Creswell1974}; \cite{Hirose1974}; \\
&&&\cite{Medcraft2012}\\    
Ethylene oxide & c-C$_2$H$_3$DO & CDMS & \cite{Muller2023_2}\\
Methanethiol &CH$_3$SH & CDMS & \cite{Zakharenko2019}\\
Glycolonitrile &HOCH$_2$CN & CDMS & \cite{Margules2017}\\ 

\bottomrule
        \end{tabular}}
        \tablefoot{For acetone, we use a corrected entry (\citealt{Ordu2019}) for the apparent issues seen in \cite{Lykke2017}.}
\end{table}

\section{Additional plots and tables}

Figure \ref{fig:model_spec} presents the fitted models of each molecule to the Band 3 data, highlighting a number of lines yet to be determined. Figures \ref{fig:n_prop} and \ref{fig:glyc} present the transitions of Ga-n-propanol and G$^\prime$Gg$^\prime$gg$^\prime$-Glycerol with the models determining their upper limits in orange. The total fit to the detected and tentatively molecules is also shown in cyan. Figure \ref{fig:Tex} presents the excitation temperatures when a measurement was possible for Band 3 results (PILS temperatures are shown for comparison). Figure \ref{fig:beams} shows the comparison of Band 3 and Band 7 beams on top of a two-dimensional Gaussian distribution representing the spatial extents of COMs (e.g., see \citealt{Jorgensen2016}). This is done to examine the difference between the mean flux in the two beams which is around 10$\%$. Table \ref{tab:non_detect} shows the upper limits found for the molecules searched for but not detected. Table \ref{tab:lines} presents the covered transitions of the (tentatively) detected molecules in the data.

\begin{figure*}
    \centering
    \includegraphics[width=14cm]{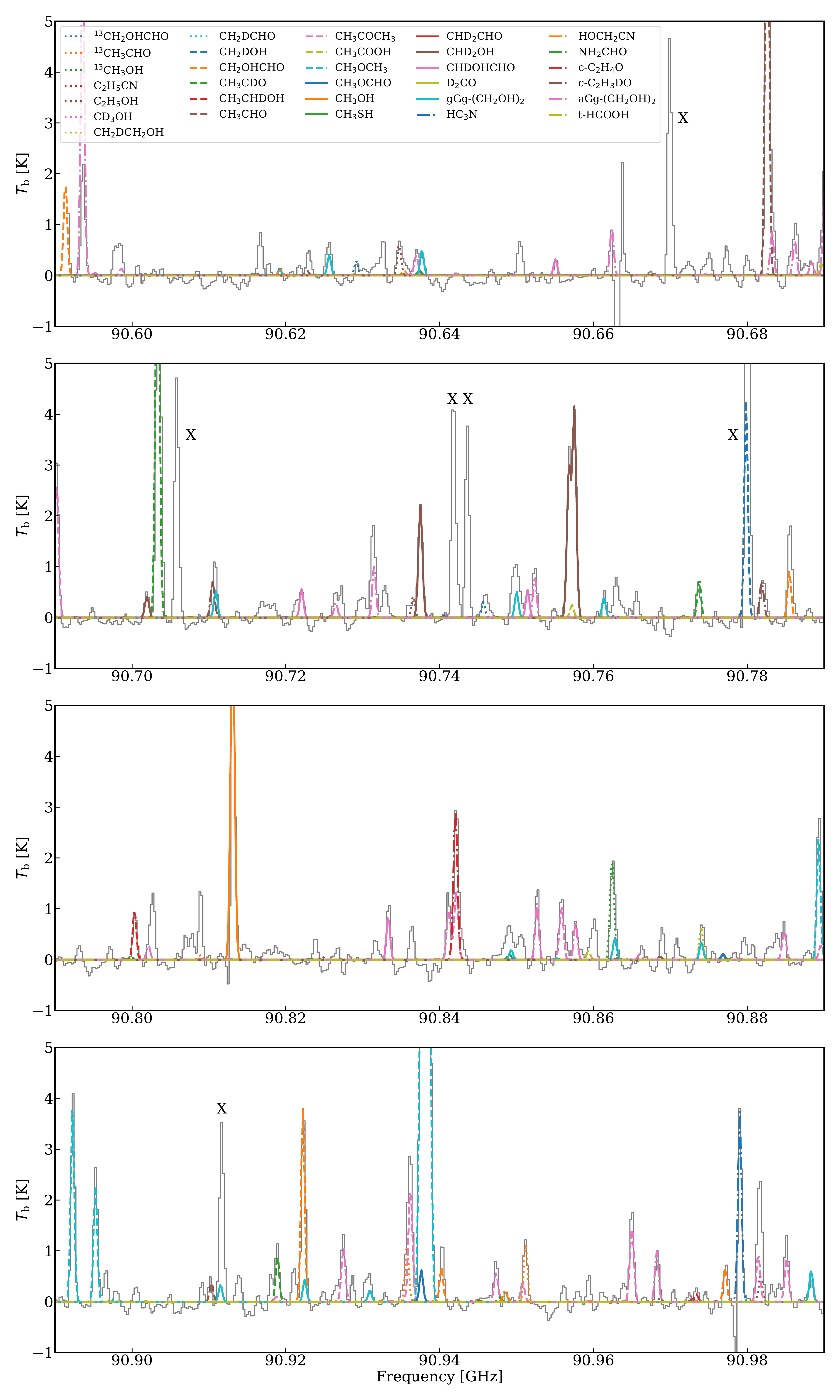}
    \caption{The fitted model for each molecule; on top of the Band 3 data in gray. For readability the y-axis limit is set to 5\,K, however, no line is overestimated except those of methanol and one line of aGg$^{\prime}$-(CH$_2$OH)$_2$, which are potentially optically thick. The lines that have an intensity higher than 2\,K (i.e., detected at a ${\gtrsim}7-10\sigma$ level), but are not identified are indicated by an `X'.} 
    \label{fig:model_spec}
\end{figure*}

\begin{figure*}
    \ContinuedFloat
    \centering
    \includegraphics[width=14cm]{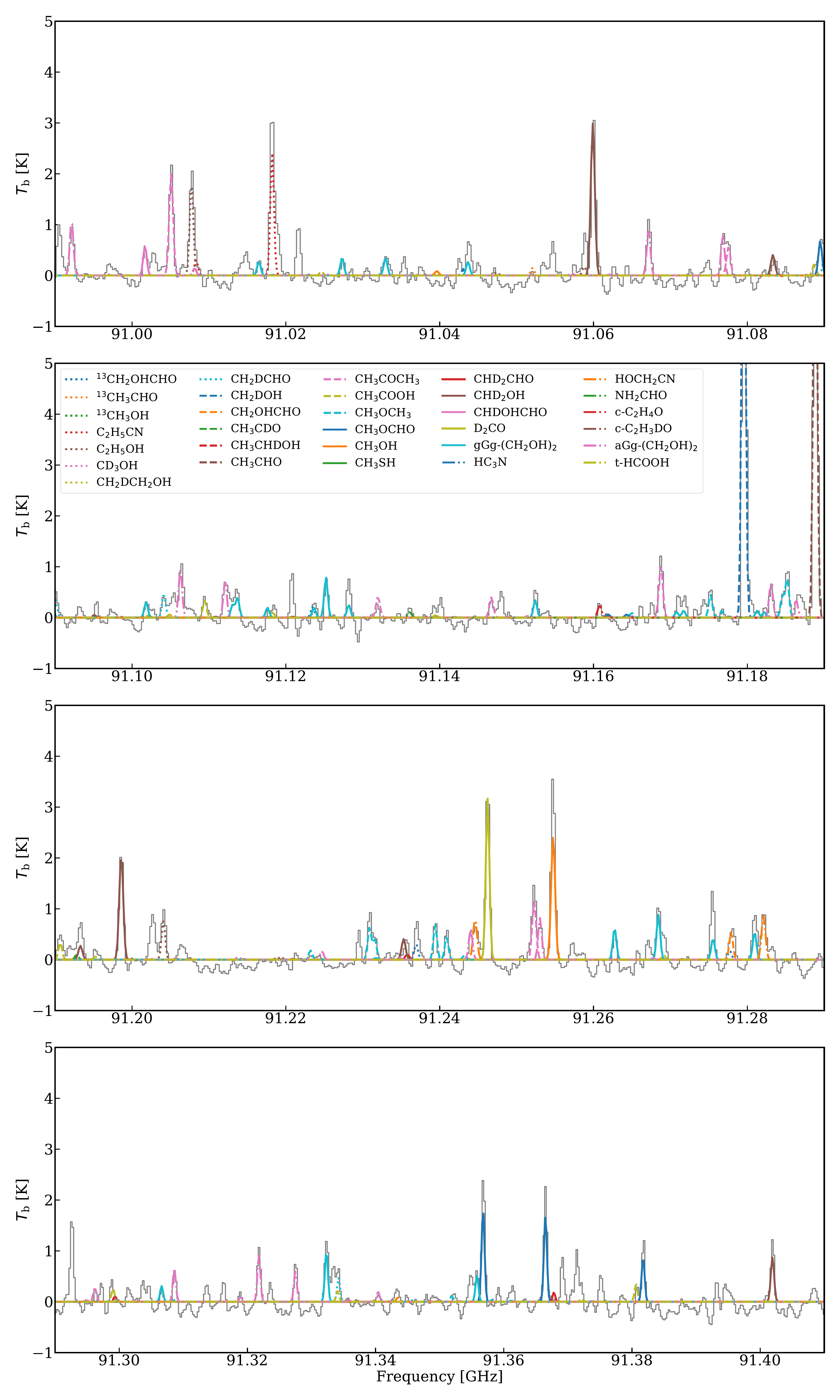}
    \caption{Continued} 
    \label{fig:model_spec}
\end{figure*}

\begin{figure*}
    \ContinuedFloat
    \centering
    \includegraphics[width=14cm]{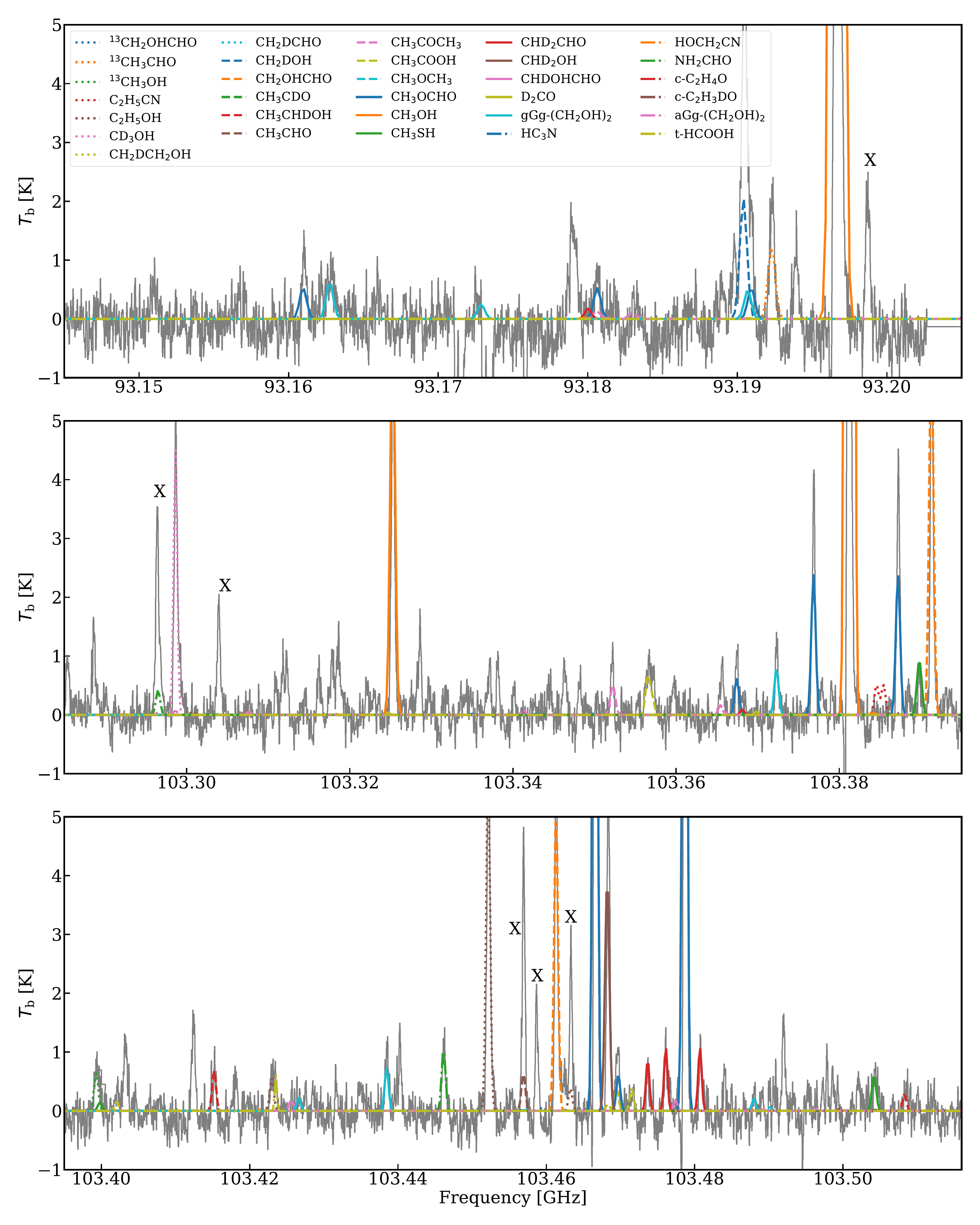}
    \caption{Continued} 
    \label{fig:model_spec}
\end{figure*}

\begin{figure*}
    \centering
    \includegraphics[width=19cm]{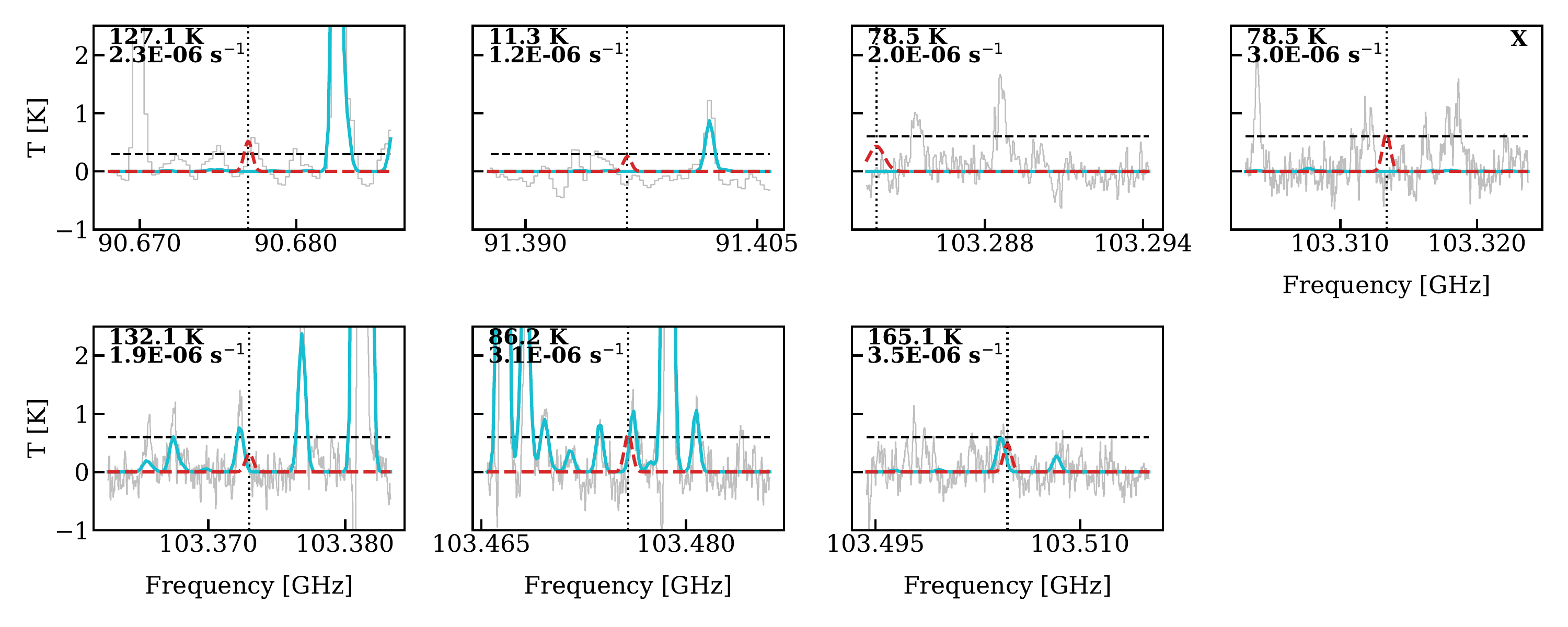}
    \caption{Lines of Ga-n-propanol and the model for its upper limit ($6\times10^{15}$\,cm$^{-2}$) in red. The symbols are the same as Fig. \ref{fig:prop}.} 
    \label{fig:n_prop}
\end{figure*}

\begin{figure*}
    \centering
    \includegraphics[width=19cm]{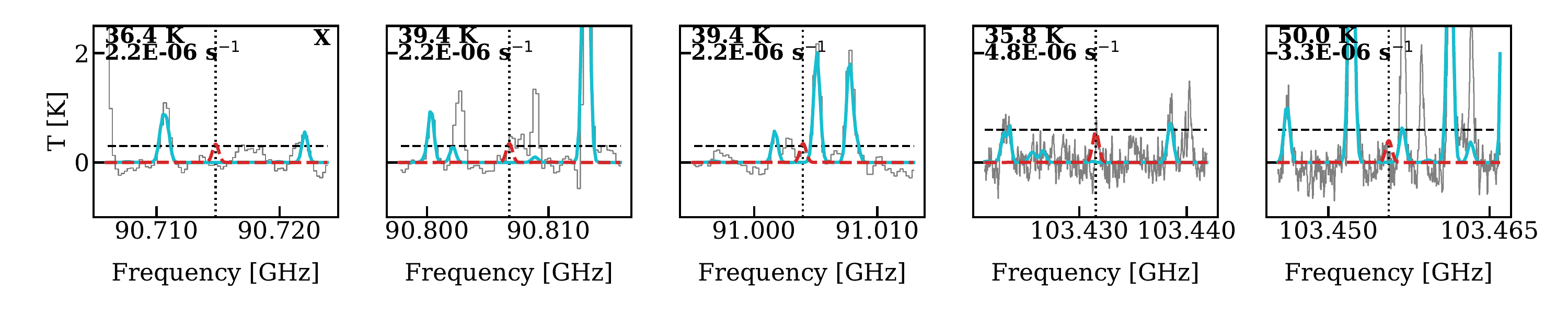}
    \caption{Lines of G$^\prime$Gg$^\prime$gg$^\prime$-Glycerol and the model for its upper limit ($7\times10^{15}$\,cm$^{-2}$) in red. The symbols are the same as Fig. \ref{fig:prop}.} 
    \label{fig:glyc}
\end{figure*}

\begin{SCfigure*}
    \centering
    \includegraphics[width=12cm]{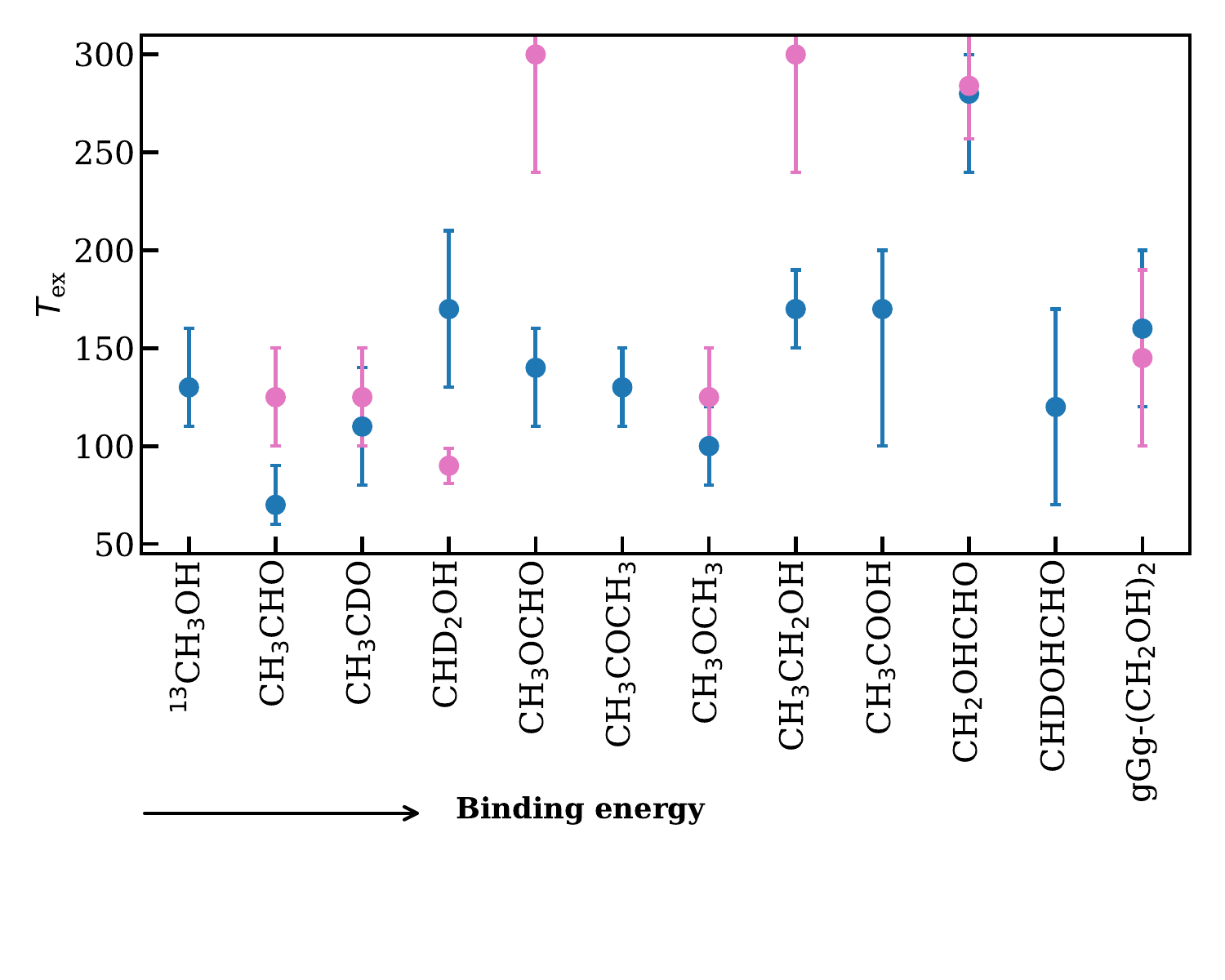}
    \caption{Excitation temperatures for molecules where a determination was possible. The species are ordered by their binding energies taken from \cite{Minissale2022} and \cite{Ligterink2023}. The binding energy of the isotopologues is assumed to be the same as the major isotopologue. Band 3 results are shown in blue and Band 7 (PILS) in pink. PILS results with a fixed temperature are not shown. The temperatures for molecules from \cite{Jorgensen2018} are shown with either $300\pm60$\,K or $125\pm25$\,K depending on the group they were associated with.} 
    \label{fig:Tex}
\end{SCfigure*}

\begin{SCfigure*}
    \centering
    \includegraphics[width=11cm]{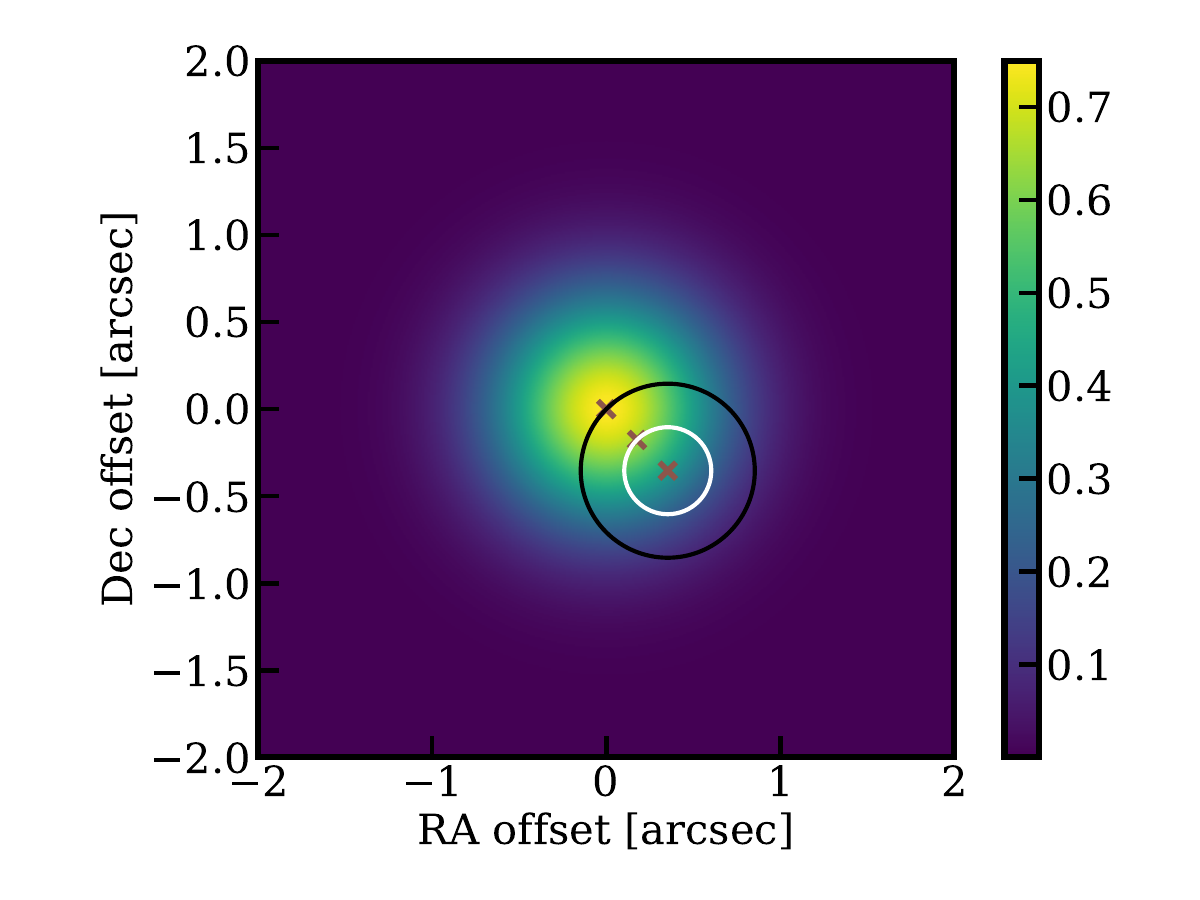}
    \caption{A two-dimensional Gaussian function with a total integration of one and a FWHM of 0.5$\arcsec$ (color scale). This is assumed as a typical emission distribution for COMs based on the results of PILS (e.g., \citealt{Jorgensen2016}). Two circular regions with radii of 0.25$\arcsec$ and 0.5$\arcsec$ are over plotted representing beams of PILS (white) and Band 3 (black), respectively. The crosses show the peak position, 0.25$\arcsec$, and 0.5$\arcsec$ offset positions.} 
    \label{fig:beams}
\end{SCfigure*}

\begin{table}
\Huge
\renewcommand{\arraystretch}{1.3}
    \caption{Upper limits of non-detections}
    \label{tab:non_detect}
    \resizebox{\columnwidth}{!}{
\begin{tabular}{@{\extracolsep{1mm}}*{3}{l}}
          \toprule
          \toprule      
        Name  & Molecular formula & Upper limit (cm$^{-2}$)\\
        \midrule     

    \hline

Methanol&CH$_3^{18}$OH& $<$5.0 $\times 10^{17}$\\
Formamide&NH$_2$CDO& $<$2.0 $\times 10^{15}$\\
Formamide&NH$_2^{13}$CHO& $<$1.2 $\times 10^{16}$\\
Formamide&cis-NHDCHO& $<$2.5 $\times 10^{15}$\\
Formamide&trans-NHDCHO& $<$5.0 $\times 10^{15}$\\
Formamide&NH$_2$CHO($v_{12}=1$)& $<$5.0 $\times 10^{17}$\\
Z-cyanomethanimine&Z-HNCHCN& $<$5.5 $\times 10^{15}$\\
\bottomrule
        \end{tabular}}
\tablefoot{Upper limits on column densities toward the 0.5\arcsec offset position from source B in a ${\sim} 1\arcsec$ beam. If a source size of 0.5$\arcsec$ was assumed, these values would increase by a factor of ${\sim}5$. The upper limits are measured by fixing $T_{\rm ex}$ to 100\,K and FWHM to 2\,km\,s$^{-1}$. Vibrational correction factors are not included in these values.}
\end{table}

\onecolumn
\begin{longtable}{llllr}
\caption{Transitions of the species studied here in the data that have $E_{\rm up} < 1000$\,K and $A_{\rm ij}>10^{-9}$\,s$^{-1}$ (not all are detected).} \\
\hline \hline
Species & Transition & Frequency & $A_{\rm ij}$ & $E_\mathrm{up}$ \\
 & J K L M & (MHz) & (s$^{-1}$) & (K) \\
\hline
\endfirsthead
\caption{continued.}\\
\hline\hline
Species & Transition & Frequency & $A_{\rm ij}$ & $E_\mathrm{up}$ \\
 & J K L M & (MHz) & (s$^{-1}$) & (K) \\
\hline
\endhead
\hline
\endfoot

CH$_3$OH  & 20 3 18 5 - 19 2 18 5 & 90\,813.078 & $2.80 \times 10^{-6}$ & 808.3\\ 
  & 20 2 18 2 - 20 2 19 1 & 91\,254.751 & $5.29 \times 10^{-8}$ & 514.3\\ 
  & 1 0 1 4 - 2 1 2 4 & 93\,196.672 & $4.19 \times 10^{-6}$ & 302.9\\ 
  & 12 3 10 1 - 13 0 13 1 & 103\,325.252 & $1.81 \times 10^{-8}$ & 228.8\\ 
  & 12 2 10 2 - 12 1 11 1 & 103\,381.258 & $3.98 \times 10^{-7}$ & 207.1\\ 
\hline 
$^{13}$CH$_3$OH  & 18 -4 15 0 - 17 -5 12 0 & 90\,862.42 & $1.17 \times 10^{-6}$ & 475.4\\ 
  & 8 -2 7 0 - 8 1 7 0 & 103\,399.296 & $6.15 \times 10^{-8}$ & 107.6\\ 
\hline 
CH$_2$DOH  & 22 3 19 0 - 21 4 17 2 & 90\,762.771 & $5.43 \times 10^{-8}$ & 580.9\\
\end{longtable}

\end{appendix}

\end{document}